| | |
|---|---|
| **Title** | **Transmission and multiple reflection mechanisms of guided streamers propagating through grounded annular electrode and interacting with grounded surface electrode** |
| **Authors** | H. Decauchy[1], T. Dufour[1] |
| **Affiliations** | [1]UMR CNRS 8256, INSERM ERL 1164, Biological Adaptation and Ageing, Institut de Biologie Paris-Seine, Sorbonne Université, Paris, France |
| **Correspondence** | Thierry Dufour, thierry.dufour@sorbonne-universite.fr |
| **Ref.** | Plasma Sources Science and Technology, Vol. 31, No. 11 (2022) |
| **DOI** | https://doi.org/10.1088/1361-6595/aca1da |
| **Summary** | The repeatable dynamics and the reversal propagation of guided streamers remains a major question of fundamental physics. In this article, trains of positive guided streamers are generated within an atmospheric pressure plasma jet supplied in helium and polarized by a high-voltage nanosecond pulse generator. The device is completed by two distant targets: a grounded annular electrode coaxially centered around the capillary through which guided streamers can propagate, and a grounded surface electrode on which they can interact. The resulting transmitted and multiple reflected guided streamers are measured combining optical characterization (fast ICCD imaging) and electrical characterization (high voltage probe and current monitors). While the electrical approach provides information on the capacitive/conductive nature of the current peaks as well as on their positive/negative value, fast ICCD imaging distinguishes whether the guided streamers are incident, reflected or transmitted. Combining these two techniques allow us to demonstrate experimentally that the reflected streamers are negative contrarily to the others. Besides, 4 types of reflections have been highlighted: a reflection (r) at the outlet of the capillary, a reflection on the grounded surface electrode (R) and two reflections (r' and r'') observed when an incident guided streamer passes through the grounded annular electrode. The two techniques agree that the characteristic propagation times are always shorter for reflected negative streamers than for the positive ones propagating forward. Hence, for a grounded annular electrode placed 3 cm away from the high voltage electrode, propagation time is 80 ns for reflection versus 250 ns for transmission. These characteristic propagation times are even shorter when the annular electrode is brought closer to the surface electrode with velocities typically higher than 300 km/s. In addition, the intensity ratios of reflected/incident guided currents drop sharply, typically losing one decade over a counter-propagation length of only 3-5 cm. Finally, all these experimental data are utilized to build an equivalent electrical model that allow to better understand the dynamics of the guided streamers and explain their transmission and reflection modes upon their interaction with the two distant grounded electrodes. |
| **Keywords** | Guided streamers, streamer counter-propagation, return stroke, streamer-magnetic field interaction |

# I. Introduction

## I.1 Preamble

Atmospheric pressure plasma jet (APPJ) devices are studied for several decades in a wide range of applications, including materials, effluents valorization and more recently life sciences [1]. This wide variety of applications has stimulated the emergence of a large variety of APPJ configurations combined to different electrical excitation modes detailed hereafter.

## I.2 Plasma applications

In the field of materials and surface science, APPJ can be used for the etching of materials like Kapton, silicon dioxide, tantalum or tungsten [2] but also for the deposition of amorphous carbon films [3] as well as composite thin films like $In_xO_y$ and $SnO_x$ [4]. APPJ are also relevant for effluents processing, especially once they are mounted in parallel to treat larger volumes. Such configurations can be used to treat textile wastewaters and more specifically dyes that are among the most complicated environmental pollutants to treat owing to their complex structures, chemical properties and molecular weights [5]. $CO_2$ conversion by APPJ is also a flagship application that paves the way for sustainable and low-carbon processes [6]. As a third field of application, APPJ's are extensively investigated to address various life science issues. Plasma jets supplied with helium or argon can generate reactive species that play an essential role to inactivate *Staphylococcus aureus* and *Escherichia coli* and more generally to decontaminate surfaces from a large spectrum of microorganisms (bacteria, bacilli, fungi, viruses) [7]. APPJ are also used in agriculture to improve seeds' germinative properties. Although they can only treat small batches of seeds, APPJ can be used to easily test a wide range of experimental conditions and then innovate larger-scale plasma processes. In medical research, APPJ's have established in many areas, especially for wound healing to trigger biological effects upon disinfection, proliferation, angiogenesis, cell migration and re-epithelialization [8]. APPJ have been successfully applied in oncology to induce antitumor effects upon *in vivo* campaigns, either following a direct approach (tumor cells exposed to plasma) [9] or an indirect approach (utilization of a plasma-activated liquid) to generate long lifespan reactive species [10].

## I.3. APPJ configurations

The diversity of the aforementioned applications and the specific know-how of each laboratory have led to the emergence of a large panel of APPJ devices. As sketched in Figure 1, it is convenient to classify APPJ's into three categories. The first includes the devices where the plasma is in contact with no electrode. Such APPJ present double dielectric barriers as sketched in the (a) configuration where two outer rings surround a dielectric capillary, in the (b) configuration where a pin electrode is embedded in a dielectric material or in the (c) configuration where the polarized electrode and the counter-electrode are wounded around a tube following a double helix structure. Such dielectric-embedded electrodes are particularly relevant for processing liquids or corrosive gases [11]. The second category corresponds to the APPJ's with single-dielectric barrier so that plasma is in contact with only one electrode. This is the case of the (d) configuration corresponding to the plasma gun device successfully applied in several medical applications [12] as well as for the (e), (f) and (g) configurations, the latter one being drastically different from the previous ones in that its counter-electrode is not in contact with the dielectric tube. The third category gathers APPJ devices where the plasma is in contact with at least one electrode and where the concept of a dielectric barrier is no more relevant. In this category, a metal surface is often used as the counter-electrode, as sketched in the (h) configuration. A variant corresponds to the (i) configuration where no counter-electrode is present. In the two latter configurations, the main role of the dielectric tube is to flush the gas flow in a preferred direction, with the ability to influence the discharge inception and its spatio-temporal dynamics [13]. However, the dielectric property of the tube cannot be utilized to prevent arc transition. For this reason, small modification of interelectrode distance or applied power can easily change the plasma from cold to thermal state.

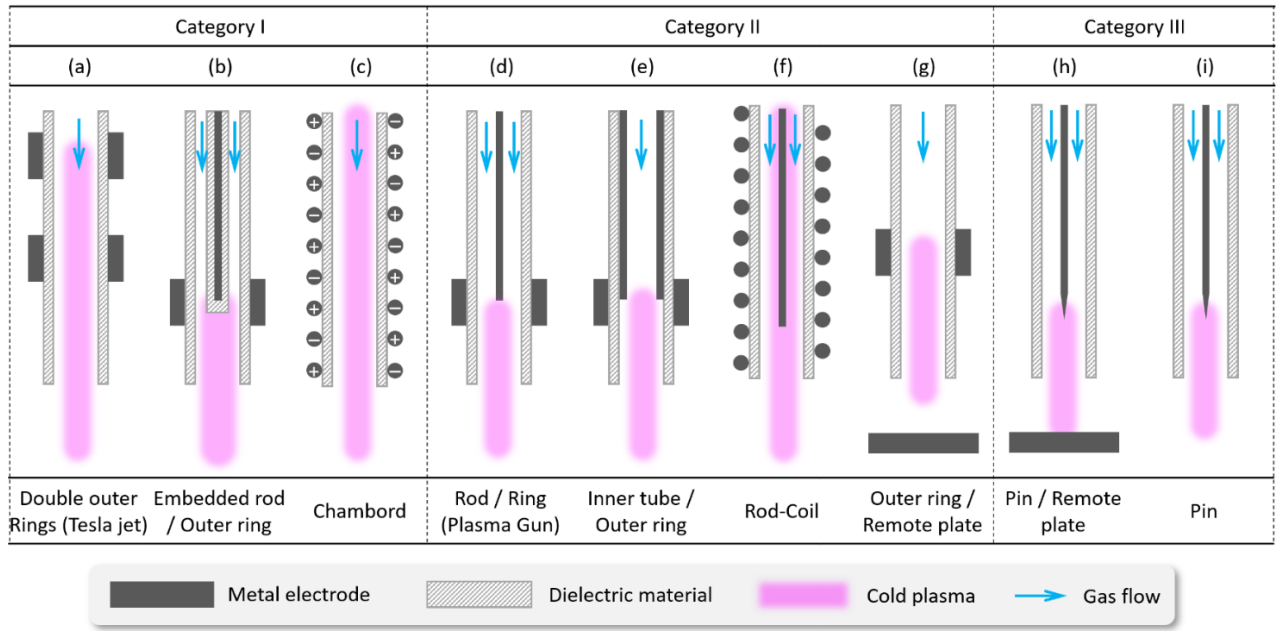

| | Category I | | | Category II | | | | Category III | |
|---|---|---|---|---|---|---|---|---|---|
| | (a) | (b) | (c) | (d) | (e) | (f) | (g) | (h) | (i) |
| | Double outer Rings (Tesla jet) | Embedded rod / Outer ring | Chambord | Rod / Ring (Plasma Gun) | Inner tube / Outer ring | Rod-Coil | Outer ring / Remote plate | Pin / Remote plate | Pin |

▮ Metal electrode ▨ Dielectric material ▮ Cold plasma → Gas flow

*Figure 1. Non-exhaustive nomenclature of AAPJ devices and their cross-sectional representations*

## I.4. Structuring of a streamer

Whatever the APPJ configuration, the physical mechanisms responsible for the ignition of a cold plasma discharge remain the same. First, the background radiation (cosmic radiation and environmental radioactivity) generates seed electrons within the gas confined in the interelectrode space. There, the applied voltage creates an external electric field that strongly accelerates these electrons, making them collide at high frequency with neutral species (atoms and molecules) and leading to their excitation and/or ionization. Primary electrons can hence be generated and accelerated by the external electric field to collide with neutrals and lead to a chain reaction commonly called "electron avalanche". As soon as the Raether-Meek criterion is met, i.e. the number of electrons in the avalanche's head is higher than $10^8$-$10^9$, the avalanche turns into a streamer: an ionization wave that propagates longitudinally and that transports electrical charges and radiative species over long distances [14]. Streamers can be detected by fast ICCD imaging and/or by electrical probing. Combining these diagnostics with simulation tools allow to model a streamer as a multi-stage structure including:

- The pre-head region which is located far upstream of the streamer's head and where seed electrons can be created through several physical mechanisms: local electric field induced by space charges, background ionization effects and photo-ionization [15], [16]. The efficiency of this latter mechanism strongly depends on the presence of both nitrogen and oxygen: while it dominates streamer propagation in air for repetition frequencies of at least 1 kHz. It depends also on the repetition frequency in the case of $N_2$ with 1 ppm $O_2$ [17].

- The head which can take the appearance of a highly emissive and small bullet (e.g. positive guided streamer generated in helium before interacting with a grounded metal target [18]) or on the contrary appear with the same optical emission of the tail, hence the denomination of filament (e.g. positive streamer discharges in $N_2$-$O_2$ gas mixtures at low pressure [17], [19].
- The tail which has the appearance of a long drag of weak intensity and decreasing as one moves away from the head. The tail can contain positive and negative charged species although its whole and macroscopic electrical charge is zero [14] [18].

The term "plasma bullet", although widely used in the literature, is somewhat misleading in that it suggests a small volume of plasma (here the head) propagating as a projectile completely independent of the device that generated it. In the case of a plasma gun, this terminology would imply that the "plasma bullet" is disconnected from the polarized electrode and therefore that the tail does not exist, which is obviously false [20], [21]. Their dynamics has been extensively investigated as part of simulation works which concluded that "plasma bullet" propagation is similar to cathode streamer propagation, i.e. ionization waves guided by the jet of flowing gas [21]. For the sake of clarity, this terminology is abandoned in this article.

## I.5. Streamers – Propagation modes

Depending on the APPJ configurations and the profile of the applied voltage, different electrical excitation modes can be reached, driving either to trains of guided streamers or non-guided streamers. A train of guided streamers is the succession of streamers periodically repeated in time and space, i.e. each streamer following the same spatial path arrangement as its predecessors. Such streamers always propagate at the same velocity (speed variations < a few percent), after a same delay time (uncertainties < a few nanoseconds) [22]. Conversely, a train of unguided streamers corresponds to a succession of streamers that are randomly distributed in time and/or in space. Thus, streamers that are repeated in time but whose propagation paths change, belong to this category, as well as those which appear at different times after the instant when the breakdown voltage is reached and occupy a different spatial arrangement at each shot [23]. As underlined by Zeng *et al.*, such streamers propagate in different directions with a typical variation of the propagation velocity of approximately 20%-50% [24]. The stochastic behavior of unguided streamers may result from their ignition jitter and from the variation of their propagation velocity from pulse to pulse [22]. The key physical parameter that differentiates guided streamers from unguided is the seed electron density that remains in the propagation channel between two successive streamers. If its value is higher than $10^9$ cm$^{-3}$, the dynamics of the plasma plume transits from a stochastic mode to a repeatable mode [22]. The unguided/guided streamers transition can also depend on the geometry and size of the capillary which impacts on the flow regimes (laminar, turbulent) and related timescales like Kolmogorov microscale. To study this transition in a fine and reliable way, it is therefore recommended to compare several rapid characterization diagnostics, for example electrical vs. optical as in this article or optical emission vs. laser induced fluorescence, as in Iseni et al. [25].

## I.6. Electrical excitation modes

The Figure 2 shows three modes of electrical excitation (DC, sine and pulses) considering the (g) and (h) configurations of Figure 1. The DC excitation mode is undoubtedly the most versatile since it permits to generate guided or unguided streamers, mostly depending on the value of the interelectrode gap. As sketched in Figure 2a, transient streamers propagate randomly at high speed from the pin electrode. They can give rise to corona streamers (Figure 2b) which vanish before reaching the counter-electrode [27]. As a result, the time profile of the discharge current is pulsed at a steady repetition frequency (a few kHz), hence testifying that the streamers are repeated in time although not in space [26]. By reducing the interelectrode gap, the avalanches can bridge the pin electrode to the grounded metal plate by forming a single filament, as sketched in Figure 2c. Then, a shortening of the inter-electrode gap leads to a discharge current that is still pulsed but with a higher frequency and smaller pulse width, hence resulting into a transient glow, as sketched in Figure 2d. This discharge is composed of time-repeated streamers and presents locally a dark space (DS) region close to the counter-electrode. For shorter interelectrode gaps, the Figure 2e indicates that the current shows a constant time profile, meaning that the discharge current is no longer carried by streamers. Shorter gap values drive to the appearance of a spark that can be either directive or branched, as sketched by the Figures 2f and 2g respectively [28]. The streamers can be considered as guided only for the directive spark because they show a highly reproducible repetition frequency and they occupy a unique spatial pattern [29]. If the power supply provides a current of sufficiently high magnitude, then the spark turns into an arc (thermal plasma). The Figure 2 proposes two other electrical excitation modes from the same "pin / remote plate" configuration. In the case of a high voltage sinusoidal polarization (Figure 2h), the discharge current shows two components: (i) a dielectric one corresponding to a sinusoidal waveform and (ii) a plasma component composed of a stochastic streamers distribution which therefore corresponds to non-guided streamers. Conversely and as illustrated in Figure 2i, guided streamers can be obtained using the same APPJ device if it is powered by high voltage pulses – either positive, negative or alternatively positive/negative – with widths that must be short enough (< 100 μs). Such guided streamers typically propagate at velocities in the range $10^2$-$10^5$ km.s$^{-1}$ [30] [31].

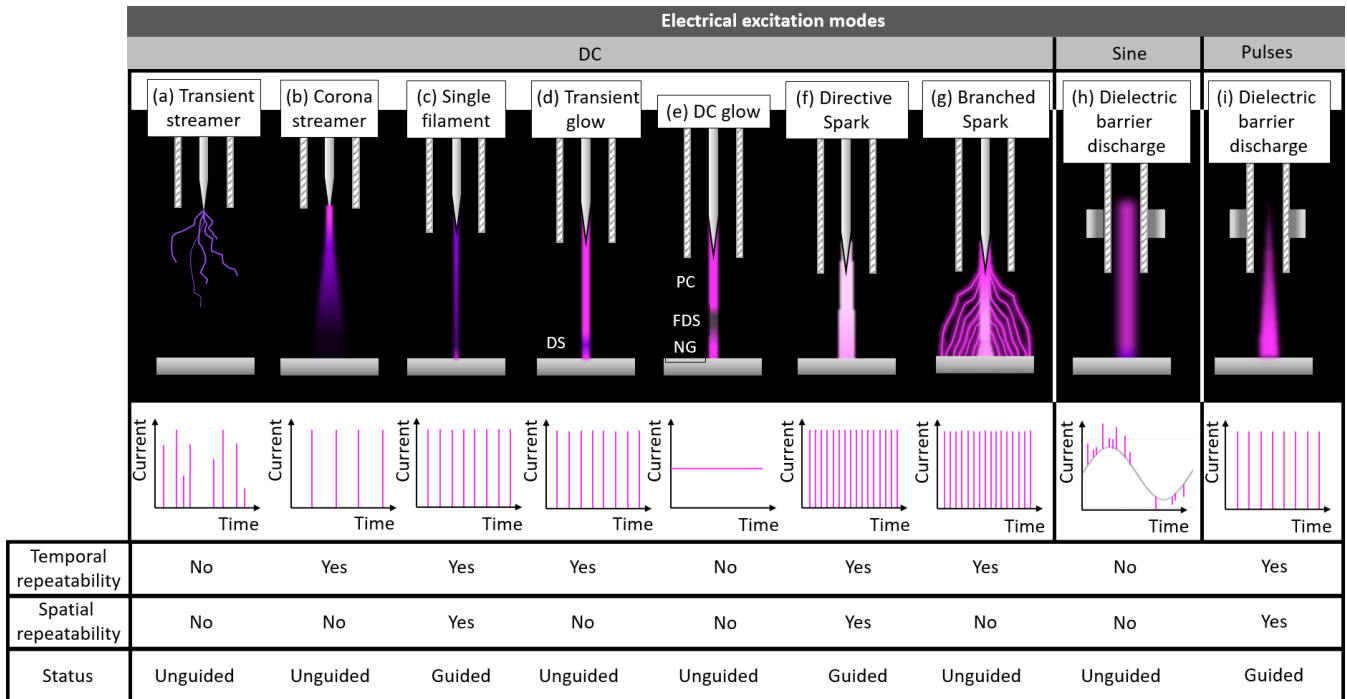

*Figure 2. Schematics of non-equilibrium atmospheric plasma discharges obtained for different electrical excitation modes and characterized either by guided streamers (temporal and spatial repeatability), non-guided streamers (random distribution of current peaks in space and/or in time) or the absence of streamers (continuous current).*

## I.7. Free mode, contact mode, propagation and reverse propagation

A plasma jet can operate either in free mode (propagation beyond the device's interelectrode gap and complete energetical dissipation in the gaseous environment) or in contact mode (propagation in the gaseous environment followed by the interaction with a condensed material – whether liquid or solid – commonly called "target"). The key parameters of a target are its dielectric permittivity and electrical conductivity which both determinate its capacitance and resistance. In turn, these two parameters control the time constant to charge the target surface.

When the streamer impinges a target of both low permittivity and conductivity (e.g. glass), an in-depth penetration of the electric field is evidenced with negligible lateral gradients [32]. As a result, the surface is charged by a fast accumulation and spreading of the streamers [33]. However, when the streamer impinges a target of both high permittivity and conductivity (e.g. metal), the electrical charging of the target is either inexistant or operates at very low velocity. Besides, the electric field does not enter the target, but a higher voltage drop remains in the gap instead (between capillary's outlet and target). As a result, a conductive ionized channel is established, bridging the polarized electrode with the target (source of electrons). There, the localized electric field immediately creates a strong discharge to rebuild the conductive channel a few hundred ns after. As a result, a backward streamer propagates inside this channel accompanied by a long-lasting diffuse discharge [32] [34].

This counter-propagation has already been characterized by Darny et al. through metastable helium density measurements (laser absorption spectrometry) and electric field distribution measurements (Poeckels probe) [35]. These two physical parameters enable the authors to explain counter-propagation as the result of an impedance mismatch between the target and the polarized electrode. This phenomenon has also been the subject of numerical simulations, notably those of Viegas et al. using a 2D axisymmetric fluid model based on drift-diffusion-reaction equations for charged species, reaction equations for neutral species and the Poisson's equation [36]. Their simulations have demonstrated that the existence of an ionized channel is necessary for the propagation of the streamers from the target to the polarized electrode. They have also shown that chemical reactions stay in the plasma plume during the 1 µs pulse if the metal target is grounded while it decreases as low as a few hundred ns if the same target is at a floating potential [36]. Interestingly, the computational investigations of Babaeva et al. reveal that for a metal target, a backward streamer could change its direction and transform into a secondary forward streamer [34]. The velocities of the backward streamer and the secondary forward streamer are higher than that of the primary forward streamer because they both propagate along an already ionized channel. Multiple forward and backward streamers reflections have also been observed and analyzed by GREMI laboratory [35].

The dielectric permittivity of a target is a fundamental question that has also strong spinoffs in plasma applications, whether in the field of materials or life sciences. As demonstrated by Yonemori et al., the density of oxygen radicals generated by an APPJ can be twice higher in the vicinity of a glass surface rather than a biological tissue (e.g. *ex vivo* skin of a murine model) [37]. For particular applications in oncology, it also appears that the plasma jet treatment can induce effects contrary to those expected if it is placed in "contact mode" with the tumor tissue, whereas the toxicity effects are considerably reduced in "remote mode" [38]. Furthermore, researchers can take many advantages from targets in medical applications: (i) they can be used to verify the absence of electrical and thermal hazards before applying plasma on preclinical models, e.g. mice, rats and pigs, (ii) they can be engineered so as to mimic the electrical response of human bodies while taking into account biophysical factors (e.g. pregnant woman, child with clammy skin, old man with metal prothesis, etc.) to customize the plasma therapy [39], [40].

# II. Experimental setup & Methods

## II.1. Plasma gun device

Streamers are generated using a plasma gun: an APPJ device composed of a quartz capillary, an inner rod electrode and an external ring counter-electrode, as sketched in Figure 3. The quartz capillary is 150 mm long with inner and outer diameters of 2.0 and 4.0mm respectively. The inner rod electrode (50 mm in length, 2.0 mm in diameter) is biased to the high voltage power supply and is called "high voltage" (HV) electrode while the outer ring counter-electrode (10 mm in length) is grounded. The central coordinate of the ring electrode ($x_{ring}$) is aligned to the end of the inner rod electrode ($x_{rod}$) so that $x_{ring} = x_{rod}$ as sketched in Figure 3. The plasma gun is supplied with helium at 1000 sccm and polarized by positive pulses of high voltage generated by a pulse generator (RLC electronic Company, NanoGen1 model) coupled with a DC high voltage power supply (Spellman company, SLM 10 kV 1200W model). In all the experiments, the high voltage magnitude is 8500 V, the duty cycle is 10 % and the repetition frequency is 5 kHz.

This plasma source is completed by two distant grounded electrodes: a grounded annular electrode (GAEL) which correspond to the external housing of the current monitor $CM_1$ and a grounded surface electrode (GSEL) that is placed 15 mm from the outlet of the plasma gun.

## II.2. Electrical measurements & Signal processing

### II.2.1. Electrical probes

Electrical parameters are measured using an analog oscilloscope (Wavesurfer 3054) from Teledyne Lecroy coupled with a high voltage probe (Tektronix P6015A 1000:1; Teledyne LeCroy PPE 20 kV 1000:1, Teledyne LeCroy PP020 10:1) and two current monitors (Pearson, 2877).

Each current monitor (CM) consists of a 16 mm thick hollow cylindrical housing, with internal and external diameters of 6 mm and 26 mm respectively. Since this housing is in metal and connected to the ground, it corresponds to a grounded annular electrode (GAEL). It contains a ferromagnetic torus (FT) characterized by its inner poloidal radius ($R_{in}$ = 7 mm), outer poloidal radius ($R_{out}$ = 13 mm) and its rectangular section (length $L_1$ = 6 mm × width $w_1$ = 12 mm). A metal wire is wound around this torus, forming N = 50 turns equidistant from each other. As sketched in Figure 3, FT is inside GAEL, the whole forming CM. In this article, while $CM_2$ is only used to measure current, $CM_1$ plays two roles: (i) being a grounded electrode thanks to its metal housing that is connected to the ground of the experimental room and (ii) measuring current thanks to its internal ferromagnetic torus. When a guided streamer passes through CM, it generates circular magnetic field lines that are perpendicular to the streamer propagation. These lines create an induced current in the ferromagnetic torus which – after calibration procedure – corresponds to the electrical current of the streamer.

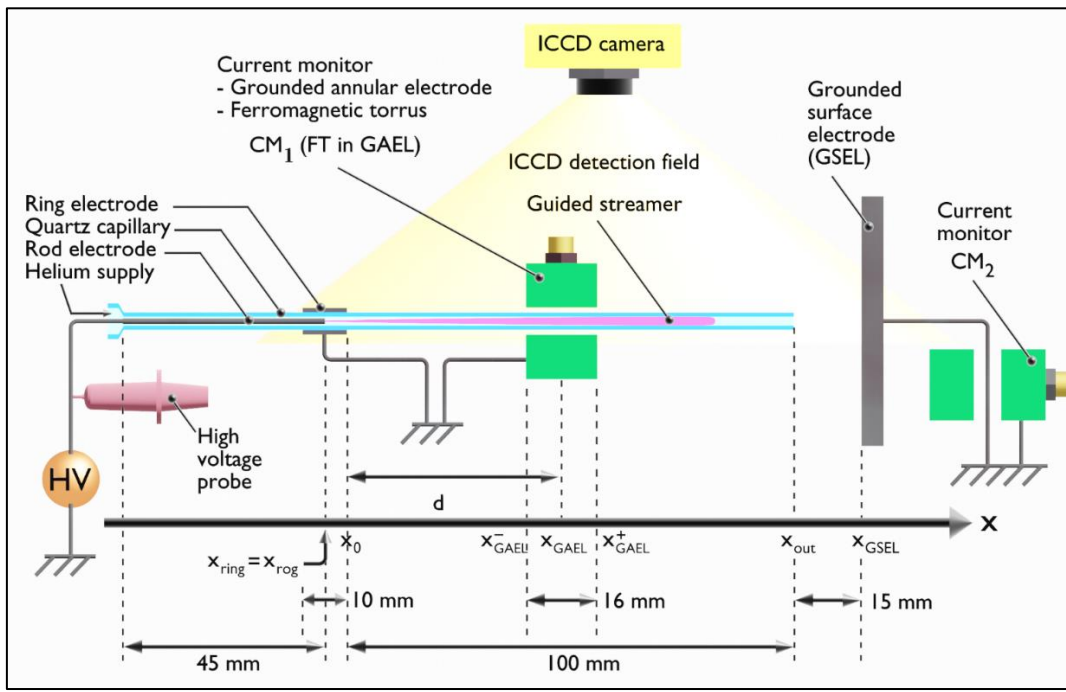

*Figure 3. Experimental setup of the plasma gun device with two distant grounded electrodes: a grounded surface electrode (GSEL) and a grounded annular electrode (GAEL) that is coaxially centered and which corresponds to the external housing of a current monitor ($CM_1$). Incident, reflected and transmitted streamers are analyzed using current monitors $CM_1$ and $CM_2$, as well as fast ICCD imaging.*

### II.2.2. High voltage pulses

An ideal signal pulse is characterized by a rising edge with a characteristic rising time $\tau_{rise} = 0$ s, a falling edge with a characteristic falling time $\tau_{fall} = 0$ s and a time width along which its amplitude is constant, as sketched in Figure 4a. In this work, the high voltage positive pulses delivered by the Nanogen power supply are close to ideal rectangular pulses: they present a magnitude of 8.5 kV, a droop of only 200V (Figure 4b), an overshoot of 13.2 kV (Figure 4e) while $\tau_{rise}$ and $\tau_{fall}$ are both equal to only 36 ns (Figures 4e and 4f). The values of the overshoot and ringing are low enough to confirm that the damping of the electronic circuit is of completely satisfactory quality. Besides, each pulse can be decomposed into a weighted summation of series of sine waves, as shown in the amplitude spectrum of Figure 4d where the signal fluctuations (mainly present during $\tau_{rise}$ and $\tau_{fall}$) see their amplitude sharply decrease with frequency. Finally, the derivative of the voltage pulse in Figure 4c results into two peaks whose magnitude multiplied by the device capacitance provides the positive and negative capacitive peaks specific to the plasma gun. These peaks are extrinsic to physical properties of the streamers (or plasma) and can therefore be considered as benchmarks for comparing streamers dynamics.

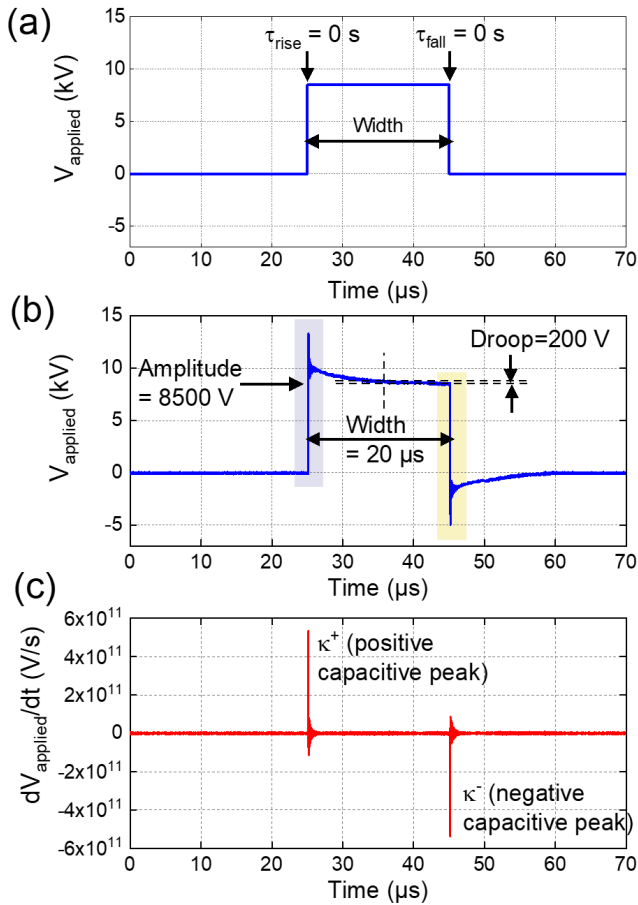

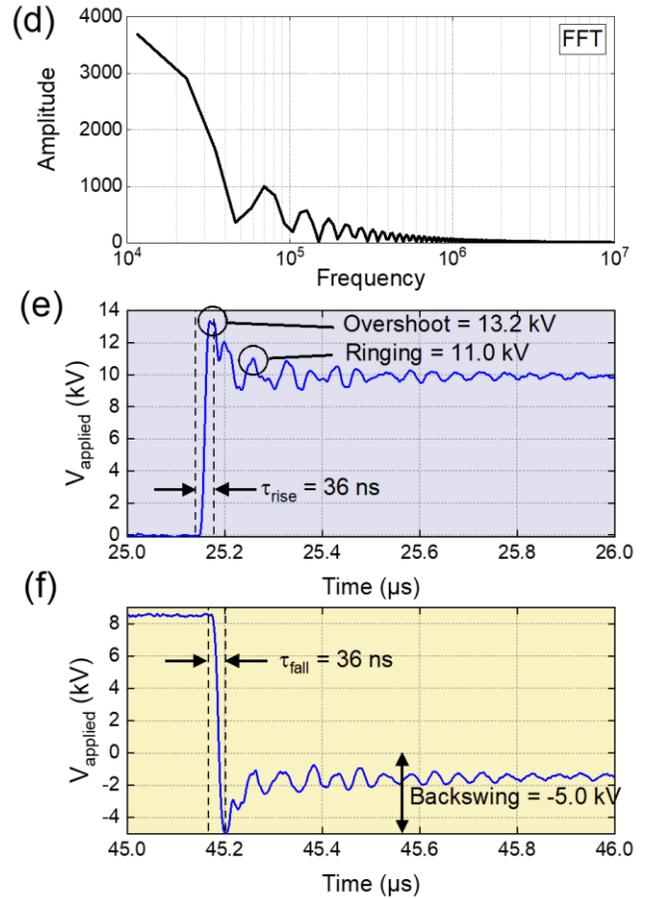

*Figure 4. (a) Ideal positive square pulse obtained with the HV power supply, (b) Real positive square pulse obtained with the HV power supply, (c) Derivative of the real pulse to evidence the positive ($\kappa^+$) and negative capacitive peaks ($\kappa^-$), (d) Fast Fourier Transform of the real pulse, (e) Enlarged view of the pulse rising edge to evidence $\tau_{rise}$, overshoot and ringing, (f) Enlarged view of the pulse falling edge to evidence $\tau_{fall}$ and backswing.*

### II.2.3. Current peaks

Electromagnetic radiation (EMR) corresponds to waves of electric and magnetic energy moving together in space. The HV power supply utilized to generate cold plasma (or streamers) is an EMR source owing to its transformers, its electrical wires which behave as transmission lines and the rising time (typically 36 ns from 0 to 10 kV). Although this EMR emission is non-ionizing – and therefore cannot remove the electrons from the atoms through space – its strength is high enough to interfere with the normal operation of transistors in a radius of 1.5 meters, and therefore with all the surrounding electronic devices, especially laptops (e.g. locking of keyboard and mouse) [41]. Since EMR can also scramble with the measurements of current performed by $CM_1$, it must be removed through an appropriate calibration procedure. For this purpose, the measurement of current peaks associated with streamer propagation must be performed considering these two configurations: the "A configuration" where $CM_1$ is placed a few centimeters away from the capillary and the "B configuration" where $CM_1$ is coaxially centered with the capillary, as sketched in Figure 5.

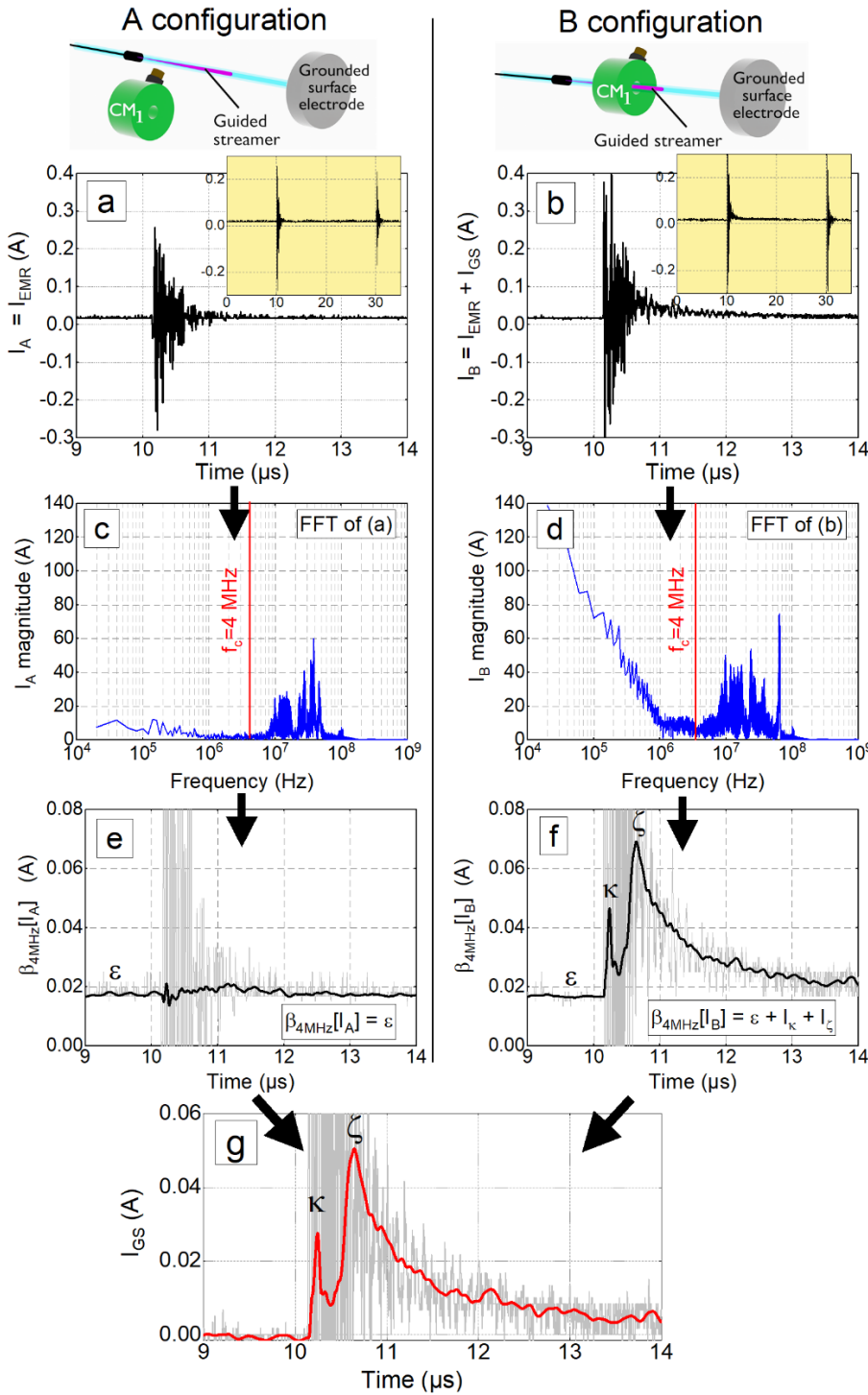

*Figure 5. Calibration procedure to measure current with CM₁ at the rising edge of a positive pulse (a) EMR current, (b) Current of guided streamer & EMR, (c) FFT of (a), (d) FFT of (b), (e) Current of (b) after Butterworth processing; (f) Current of (b) after Butterworth processing; (g) Conductive current peak of the guided streamer (ζ) and capacitive current peak of the device (κ).*

In the A configuration, CM₁ can only measure the current resulting from the EMR emission which is labelled $I_{EMR}$ in equation {1}. The Figure 5a shows an enlarged view of the $I_{EMR}$ profile on the rising edge of the HV pulse while the inset recalls the existence of an $I_{EMR}$ component for each side of the pulse: one on its rising edge and the other on its falling edge. In the B configuration, the total current measured by CM₁ ($I_B$) has three components as stated in equation {2}: the first corresponds to the emission of electromagnetic fields ($I_{EMR}$), the second is related to the capacitive current peak of the device ($I_κ$) and the third stands for the conductive current peak of the guided streamer ($I_ζ$). In this configuration, the Figure 5b shows the profile of $I_B$ at the rising edge of the HV positive pulse. Since its profile is too noisy to clearly distinguish the three aforementioned components, a Fast Fourier Transform (FFT) analysis can be performed to assess noise versus discrete frequency components for B configuration as well as for A configuration. As shown in Figure 5c, the FFT analysis of $I_A$ reveals that the electromagnetic interferences operate at frequencies typically higher than 4 MHz. In comparison, the Figure 5d represents the FFT spectrum of the signal in the B configuration, where the capacitive and conductive current peaks occur at a lower frequency range. A Butterworth low-pass filter of order n = 3 is applied with cut-off frequency ($f_c$) and sampling frequency ($f_s$) of 4 MHz and 2 GHz respectively. This processing is labelled $\mathcal{B}_{4MHz}$ and is characterized on a logarithmic Bode diagram by its gain at order 3 ($G_3$) which decreases linearly towards -∞, at a rate of -60 dB/decade following equation {3}. The Butterworth processing of $I_A$ leaves a residual background ε which remains constant at a value of about 18 mA, as represented in Figure 5e and expressed in equation {4}. In the B configuration, the Figure 5f shows the result of the same processing applied to $I_B$: the current intensity profile is composed of the ε current background, the capacitive current peak ($I_κ$) and the guided streamer current peak ($I_ζ$) (equation {5}). Finally, the useful components of the current intensity profile measured by CM₁ – namely $I_κ$ and $I_ζ$ – are obtained by applying Equation {6} and are unambiguously visible in Figure 5g.

$$I_A = I_{EMR} \quad \{1\}$$

$$I_B = I_{EMR} + I_\kappa + I_\zeta \quad \{2\}$$

$$G_3(f) = \cfrac{1}{\sqrt{1 + \left(\cfrac{f}{f_c}\right)^6}} \quad \{3\}$$

$$\mathcal{B}_{4MHz}[I_A] = \varepsilon \quad \{4\}$$

$$\mathcal{B}_{4MHz}[I_B] = \varepsilon + I_\kappa + I_\zeta \quad \{5\}$$

$$I_\kappa + I_\zeta = \mathcal{B}_{4MHz}[I_B] - \mathcal{B}_{4MHz}[I_A] \quad \{6\}$$

## II.3. Fast ICCD imaging & Signal processing

Since guided streamers are transient and low emissive phenomena, the observation of a single one of them requires very specific equipment like streak camera whose temporal resolution is close to 800 fs or even less [42] In our case, the radiative emission of the plasma jet is collected by an intensified charge-coupled device (ICCD) camera from Andor company (model Istar DH340T). It has a 2048 x 512 imaging array of 13.5 μm x 13.5 μm pixels) and an optical gate width lower than 2 ns. Although this camera is equipped with high intensification technology, its time resolution is lower than that of a streak camera, so that it is a mandatory to study not a single but a train of guided streamers. This means collecting at regular time intervals a large number of guided streamers ($N_{GS}$) and summing their emissions on a single image. The Solis software enables such operations combining the "kinetic series" acquisition mode and the "DDG" gate mode. Before explaining the procedure for creating an image, the three following points must be reminded:

(i) The HV power supply delivers pulses at the repetition frequency of 5 kHz for a duty cycle of 10%. Therefore, the width of a single pulse is 10% / 5kHz = 20 μs and the repetition period is $T_{rep}$ = 1 / 5kHz = 200 μs (See Figure 5g). As observed in the Figure 5g, a single guided streamer has a duration of few μs, approximately 3 μs. Therefore, the ICCD observations can be achieved only the first microseconds that follow the rising edge of each pulse. The remaining 20 − 3 = 17 μs can be ignored.

(ii) A kinetic series is composed of several scans, each scan comprising a given number of acquisitions. The kinetic series is defined by the 3 following parameters: the exposure time of a single scan ($\tau_{scan}$), the number of accumulations ($N_{acc}$) and the length which corresponds to the number of scans taken in the kinetic series ($L_{kin}$). Here, we consider that $\tau_{scan}$ = 5 s, $N_{acc}$ = 1 and $L_{kin}$ = 850. As a result, one can easily deduce the number of guided streamers (i.e. pulses, i.e. acquisitions) collected upon a single scan, namely $N_{GS} = \frac{\tau_{scan}}{T_{rep}} = \frac{5s}{200\mu s} = 25000$ as sketched in Figure 6. Besides, the total duration of a kinetic series can also be assessed as $\tau_{scan} \times N_{acc} \times L_{kin}$ = 4250 s = 1h11 min. Therefore, depending on whether one wishes to observe all or part of the forward / backward propagation phenomenon, the $L_{kin}$ value is carefully chosen to find the best compromise between measurement accuracy and measurement time.

(iii) The DDG mode is characterized by a gain G = 1500, a gate width w = 2 ns and a delay δ that can vary from 0 to 850 ($L_{kin}$ value) per step of 1 ns. This delay is defined with respect to

$t_{trigger}$: the instant corresponding to the appearance of the capacitive current peak (κ) (see Figure 5g). Consequently, and as shown in Figure 6 for Acq. 1, two successive gate widths present an overlap of 1 ns.

The procedure for creating an image is performed in 850 scans. During scan #001, while 25 000 pulses are carried out (or guided streamers are generated), the ICCD camera acquires only the first ns of each one (w = 2 ns) for a delay always maintained at δ = 0 ns, as shown in Figure 6. These 25 000 acquisitions are summed to constitute a unique acquisition specific to scan #001. During scan #002, 25 000 new pulses are achieved while the ICCD camera acquires only a tiny part of each one defined by w = 2 ns and a delay δ fixed at 1 ns. These 25 000 acquisitions are summed to constitute a new acquisition specific to #002. More generally, for each new scan k (with k <851), 25 000 pulses are carried out, each of them being partially captured by the ICCD camera over a time interval always set at w = 2 ns and shifted per 1ns-step so that once k = $L_{kin}$ = 850, δ has a value as high as 849 ns.

Considering that each ICCD picture is a matrix of 2000 columns of pixels by 250 rows of pixels, and that the streamer propagates along the rows, the propagation velocity of the guided streamers ($v_{GS}$) is measured as follows: (i) the values of the pixels are summed along each column so that integrated emissivity is represented in a 2000×1 matrix. The highest value of integrated emissivity can be associated with the ionization front of the guided streamers ($x_{IF}$ coordinate). As expressed in equation {7}, the velocity is measured at locations $x_{IF}$ from two consecutive ICCD pictures so that $t_{k+1} - t_k = 1ns$:

$$v_{GS}(x_k) = \frac{dx}{dt} = \frac{x_{IF}(t_{k+1}) - x_{IF}(t_k)}{t_{k+1} - t_k} \quad \{7\}$$

# III. Results

## III.1. Guided streamers interacting with a distant grounded surface electrode (GSEL)

### III.1.1. Transmission and reflection resulting from incident positive guided streamers

We propose to investigate the propagation mechanisms of guided streamers generated along the capillary of Figure 3 before reaching the grounded surface electrode (GSEL) located 15mm away from $x_{out}$. No current monitor ($CM_1$ and/or $CM_2$) is present in the experimental setup; only fast ICCD imaging is achieved, as illustrated by the photos compiled in Figure 7. The photos are taken at different time intervals between 0 ns (appearance of the capacitive current peak) and 3000 ns: a time lapse corresponding to $\frac{3\ \mu s}{20\ \mu s}$ = 15 % of the pulse width. Since the guided streamer is generated at the rising edge of the voltage pulse and propagates along the increasing x coordinates, it is called "incident guided streamer" and it carries a positive charge, hence the notation: $GS_1^+$. The analysis of streamers propagation can be achieved following the four stages:

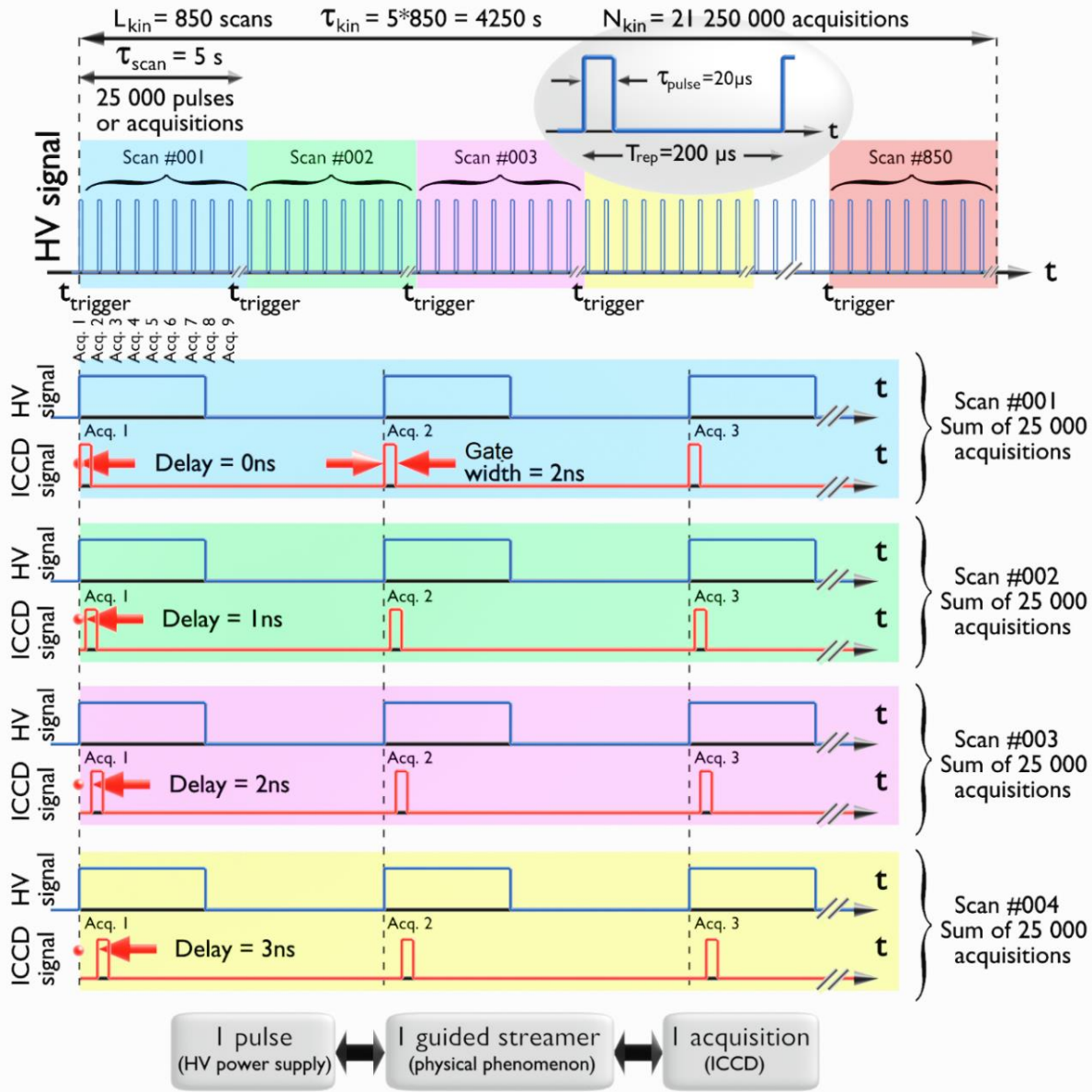

*Figure 6. Characteristic durations of the high voltage signal supplying the plasma gun and of the fast ICCD imaging signals.*

(i) From 0 to 424 ns, $GS_t^+$ propagates from the inner HV electrode to the capillary's outlet, i.e. from $x_{ring}$ to $x_{out}$, as sketched in Figure 3. Its head is clearly visible while its tail – connecting the head to the HV electrode – appears as a vanishing region only detectable in the vicinity of the head. $GS_t^+$ slows down as it nears the capillary's outlet.

(ii) From 425 ns to 469     ns, the streamer is transmitted ($GS_t$) out of the capillary to reach the grounded surface electrode after crossing an air gap of 15 mm. The head of the streamer interplays with the grounded surface electrode during almost 6 ns. Then, its optical emission vanishes while a reflected guided streamer ($GS_r$) is simultaneously forming in the air-gap.

(iii) From 470 ns to 1099 ns, a reflection (or counter-propagation) of the guided streamer is observed ($GS_r$) inside the capillary while its emissivity increases. Simultaneously, the part of $GS_r$ which remains outside the capillary does not extend anymore and its optical emission decreases. This phenomenon is in agreement with the simulations and experimental works of Babaeva et al. and Darny et al. respectively [34] [43].

(iv) From 1100 ns to 3000 ns, a second reflection is observed from GSEL: a guided streamer ($GS_R$) is emitted from GSEL to the capillary. However, $GS_R$ cannot penetrate inside the capillary and it shows a much lower emissivity than that of $GS_t^+$.

The positive or negative sign of $GS_t$, $GS_r$ and $GS_R$ cannot be determined through fast imaging but this issue is addressed in the section III.2.

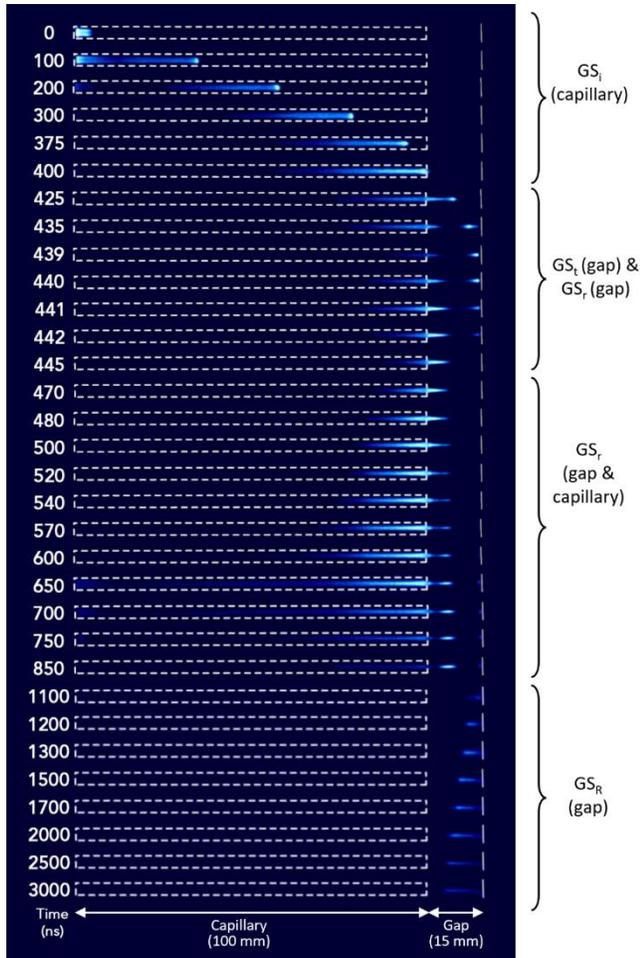

Figure 7. Photo sequence showing an incident guided streamer ($GS_i$) propagating in the capillary, then transmitted ($GS_t$) in the ambient air along a 15 mm gap before reaching the grounded surface electrode (GSEL). Two guided streamers are successfully reflected: $GS_r$ from the gap to the capillary and $GS_R$ only along the gap. Measurements achieved without $CM_1$.

In Figure 8, fast ICCD imaging is performed on the falling edges of the voltage pulses to highlight the propagation profile of the negative incident guided streamers ($GS_i^-$). Although positive and negative guided streamers have propagation velocities of the same range of order ($\approx 1 - 2.10^5\ m/s$), several discrepancies must be underlined: (i) $GS_i^-$ shows a more diffuse profile (head and tail) than that of $GS_i^+$, (ii) as evidenced by the photos at t = 100, 200 and 300 ns, $GS_i^-$ shows a highly emissive region in the immediate vicinity of the HV electrode which vanishes as the streamer propagates, (iii) $GS_i^-$ remains confined inside the capillary and never reaches GSEL, so that no reflected guided streamers can be obtained, at least in our experimental conditions. Given the time constants associated with the falling edge and the backswing, it is assumed that the highly emissive region that remains confined in the vicinity of the HV electrode is associated with the electric field reversal (falling edge) while the propagation of $GS_i^-$ is associated with the -5 kV backswing.

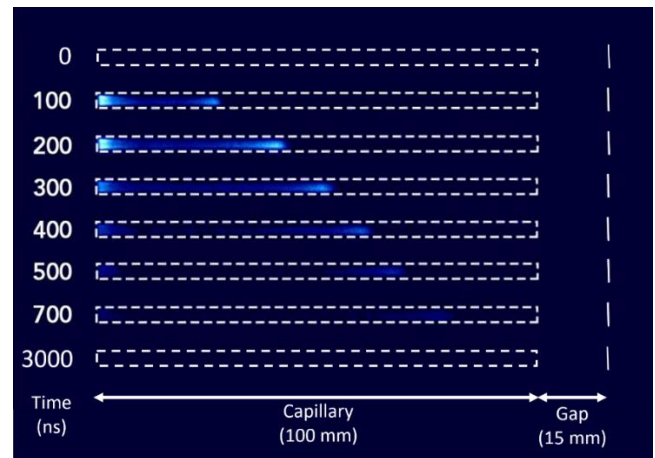

Figure 8. Photo sequence of an incident guided streamer carrying a negative charge ($GS_i^-$) and propagating along the capillary without being able to exit it. No backward propagation is observed. The $GS_i^-$ shows a highly emissive region at the immediate vicinity of the HV electrode. Measurements achieved without $CM_1$.

### III.1.2. Transmission and reflection resulting from incident negative guided streamers

When a plasma gun is supplied with a sinusoidal high voltage, the positive streamers appear on the positive half periods (or positive applied voltage) while the negative streamers appear on the negative half periods (or negative applied voltage). In the case of a plasma gun supplied with ideal positive pulses, the situation would be totally different since the high voltage would remain always positive, as sketched in Figure 4a. This means that negative guided streamers would be generated on the falling edge (Figure 4b) where the voltage remains positive but decreases to 0, hence inducing a reverse electric field. However, in our experimental study, each positive pulse presents a backswing of -5 kV, as indicated in Figure 4f which lasts over a few 100 ns. Therefore, it is expected that the negative guided streamers are the result of two components: (i) the falling edge of the positive pulse upon the first 30 ns and (ii) the -5 kV backswing upon the following hundreds of ns.

## III.2. Guided streamers interacting with two distant grounded electrodes: the annular electrode (GAEL) and the surface electrode (GSEL)

### III.2.1. Electrical characterization

We propose to study the propagation mechanisms of guided streamers as they interact with two distant grounded electrodes: the grounded annular electrode (GAEL) coaxially-centered to the capillary, followed by the grounded surface electrode (GSEL), as sketched in Figure 3. In this experimental setup, the two current monitors are present, so that the current profile of the guided streamers can be detected at two distinct locations. This profile can take the appearance of 2 or 3 current peaks among those listed in Table 1.

| Symbol | Designation |
|---|---|
| $\kappa_i^+$ | Capacitive current peak ($\kappa$) that is incident (i) with a positive charge (+) |
| $\zeta_i^+$ | Conductive current peak ($\zeta$) associated to the propagation of an incident (i) guided streamer whose electrical charge is positive (+) |
| $\zeta_t^+$ | Conductive current peak ($\zeta$) associated to the propagation of a transmitted (t) guided streamer whose electrical charge is positive (+) |
| $\zeta_r^-$ | Conductive current peak ($\zeta$) associated to the propagation of a reflected (r) guided streamer whose electrical charge is negative (–). The negative sign of this polarization is justified later in Figure 10. |

*Table 1. Types of current peaks associated to the guided streamers and that can be evidenced on the oscilloscope.*

From these peaks, five characteristic durations can be defined, as sketched in Figure 9a:

- $\tau_{\kappa\kappa}$ is the duration between the instant when the capacitive current peak is measured by CM$_1$ and the instant when the capacitive current peak is measured by CM$_2$, i.e. respectively the black and red $\kappa$ peaks in Figure 9b.
- $\tau_{\kappa\zeta}$ is the duration between the conductive current peak of a guided streamer (either $\zeta_i^+$ or $\zeta_t^+$) and its related capacitive current peak (either $\kappa_i^+$ or $\kappa_t^+$).
- $\tau_f$ is the time required by $GS_t$ to propagate forward (f) from CM$_1$ to GSEL, which is roughly the same as from CM$_1$ to CM$_2$.

- $\tau_b$ is the duration required by $GS_r$ to propagate backward (b) or counter-propagate from GSEL to GAEL, which is roughly the same as from CM$_2$ to GAEL.
- $\tau_{f+b}$ corresponds to the time required by a guided streamer to propagate (forward, f) from GAEL to GSEL and then come back to GAEL (backward, b), that is to say the time required to achieve the GAEL-GSEL-GAEL round trip.

A more analytical description of these characteristic durations is proposed in Appendix IX.1.

Now that the characteristic peaks and durations are defined, the variation of their values can be studied for different locations of GAEL with respect to the HV electrode. Based on different values of d namely 1 cm (Figure 10a), 3 cm (Figure 10b), 5 cm (Figure 10c), 7 cm (Figure 10d) and 9 cm (Figure 10e), several observations are noteworthy:

- The capacitive peak ($\kappa$) is detected by CM$_1$ at $x_{GAEL}$ and by CM$_2$ at x=$x_{GSEL}$. These peaks, whose magnitudes are directly proportional to the derivative of the applied voltage, can be considered as temporal references since their positions do not depend on the value of d, contrarily to the conductive current peaks.
- Two conductive current peaks are measured by CM$_1$ while only one is detected by CM$_2$.
- For increasing values of d, an increase in $\tau_{\kappa\zeta}$ is observed and consequently of $\tau_f$, $\tau_b$ and $\tau_{f+b}$ as well.

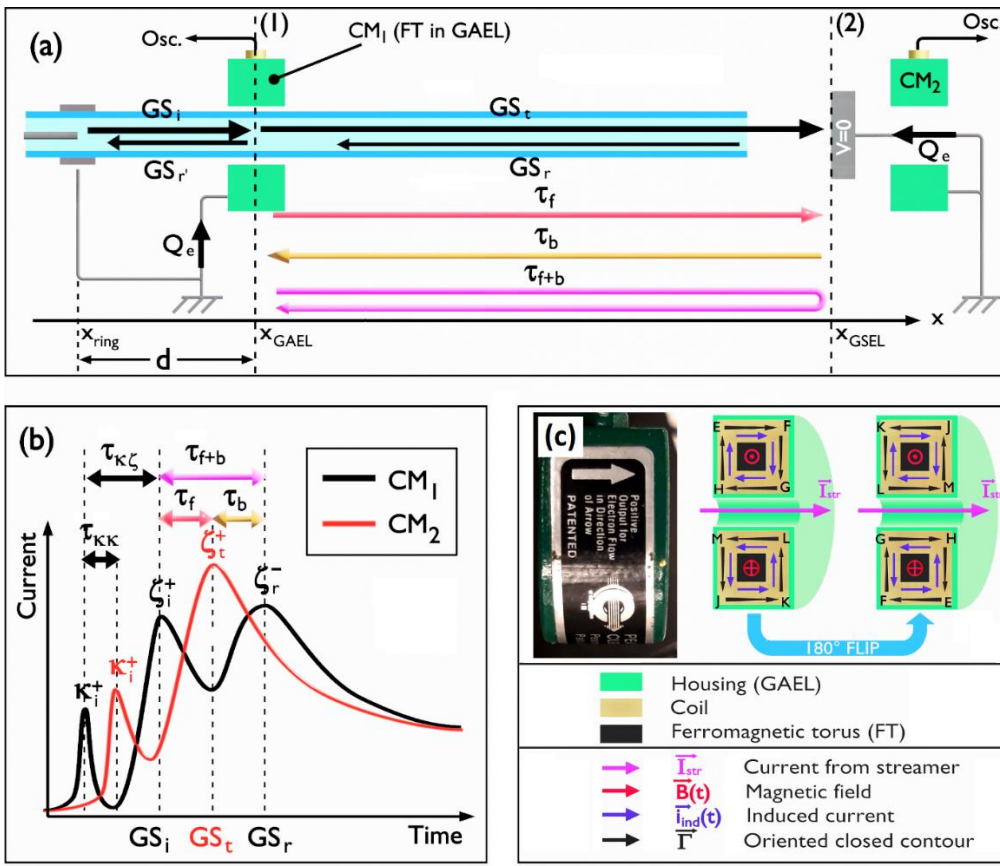

Figure 9. (a) Characteristic propagation times associated with the conductive current peaks of the incident, reflected and transmitted guided streamers, (b) Representation of the capacitive ($\kappa$) and conductive ($\zeta$) peaks associated to the guided streamers and measured upon electrical characterization to assess the aforementioned characteristic times, (c) Schematics of a current monitor (CM) including ferromagnetic torus (FT) and grounded annular housing (GAEL) explaining the sign – whether positive or negative – of the induction current.

In Figure 10, all the conductive current peaks associated with the reflected guided streamers are positive, which – at first glance – might suggest that the reflected streamers ($GS_r$) carry a positive charge ($\zeta_r^+$). However, as Figure 9c reminds us, depending on whether $CM_1$ is oriented in the conventional or opposite direction to the electron flow, it returns a positive or negative value of the same measured current ($I_{str}$). In Figure 9c, whether before or after the 180°-flip of $CM_1$, the streamers are always oriented from left to right, which means that the orientation of the magnetic field remains unchanged, as well as the induced current circulating in the coil. However, the 180°-flip has changed the orientation of the closed contour and therefore the orientation of current in input and output of the coil. For this reason, any guided streamer that counter-propagates ($GS_r$) through $CM_1$ carries a negative charge ($\zeta_r^-$) if its conduction current peak appears positive.

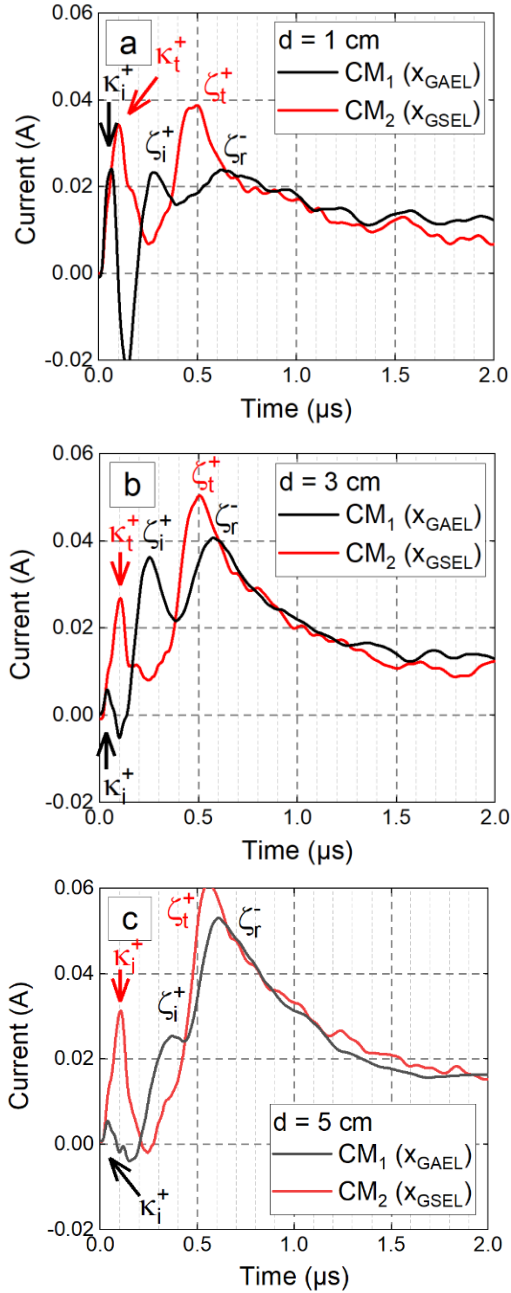

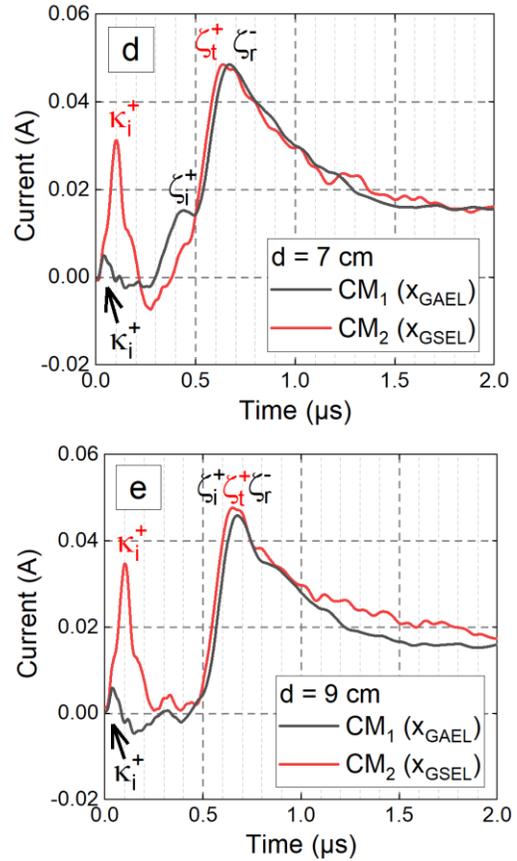

*Figure 10. Temporal profiles of current peaks associated with guided streamers and measured by $CM_1$ and $CM_2$ at $x_{GAEL}$ and $x_{GSEL}$ respectively, for d = 1, 3, 5, 7 and 9 cm ($\kappa$: capacitive current peak, $\zeta$: conductive current peak).*

### III.2.2. Relevant parameters obtained from electrical characterization

As plotted in Figure 11a, when d is increased from 1 to 9 cm, $\tau_{\kappa\kappa}$ remains roughly constant with an average value of 59.2 ns ± 2.3 ns. This means that changing d has no significant impact on the kinetics of a same capacitive peak measured at $x_{GAEL}$ and $x_{GSEL}$. The reason is that the capacitive peaks are specific to the plasma gun device and decorrelated from the streamers propagation; $\tau_{\kappa\kappa}$ can therefore be considered as a temporal benchmark to characterize the kinetics of the conductive peaks.

The variation of $\tau_{\kappa\zeta}$ versus d is represented in Figure 11b considering measurements performed by $CM_1$ (time lapse between $\kappa_i^+$ and $\zeta_i^+$) and $CM_2$ (time lapse between $\kappa_t^+$ and $\zeta_t^+$). A linear fit of the curves in Figure 11b indicates slopes with values of approximately 30 ps/cm. No datapoint is reported for $CM_1$ at d = 9 cm because the conductive current peak ($\zeta_i^+$) of the incident guided streamer ($GS_i$) is overlapped by the conductive current peak ($\zeta_r^+$) of the reflected guided streamer ($GS_r$), as evidenced in Figure 10e. Figure 11b shows that by remoting GAEL ($CM_1$) from the HV electrode, the conductive current peak ($\zeta_i^+$) associated to $GS_i$ is delayed since the capacitive current peak ($\kappa_i^+$) always appears at the same instant. The reason is that the further GAEL is from the HV electrode, the slower the incident streamer arrives before reaching GAEL. Using $CM_2$ leads to the same delay observed between $\zeta_t^+$ and $\kappa_t^+$. As a result, the time gap between these two current monitors is always close to 170 ns (see vertical arrow) whatever the value of d.

The characteristic propagation times of the guided streamers can be deduced from Figure 11c. An increase in d drives to shorter values of $\tau_f$ and $\tau_b$, as well as of $\tau_{f+b}$ while always keeping the relation $\tau_{f+b} = \tau_f + \tau_b$. The values of $\tau_f$ are always lower than those of $\tau_b$, meaning that the velocity of a guided streamer is directly correlated with its electrical charge: a positive guided streamer transmitted through GAEL propagates slower than a negative guided streamer reflected by GSEL. When d is increased, the GAEL-GSEL distance is necessarily decreased, so that the characteristic propagation time associated to the round-trip ($\tau_{f+b}$) is reduced as well.

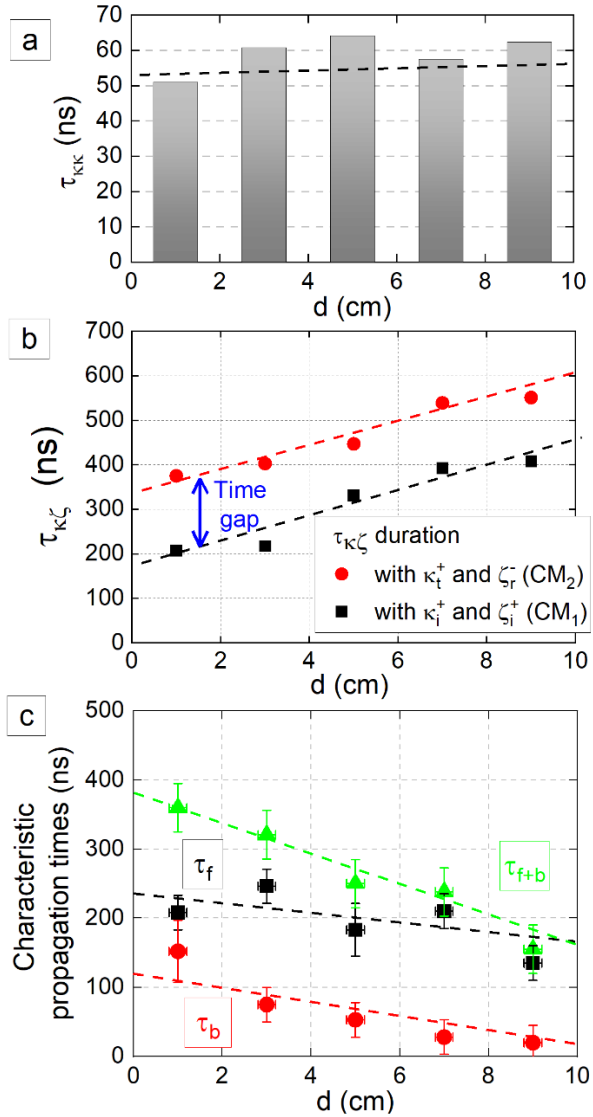

Figure 11. Influence of d on (a) $\tau_{\kappa\kappa}$ (temporal interval of a same capacitive peak measured at $x_{GAEL}$ and $x_{GSEL}$), (b) $\tau_{\kappa\zeta}$ : temporal interval between capacitive and conductive peaks versus d, (c) Characteristic propagation times of the guided streamers versus d.

### III.2.3. Fast ICCD characterization

In addition to the previous electrical characterizations, a fast ICCD Imaging study is performed to track the propagation and counter-propagation kinetics of the guided streamers. If the Figure 5 indicates that these phenomena typically occur on a time scale of a few μs (roughly 13 μs – 10 μs = 3 μs), the ICCD results obtained in Figure 7 demonstrate that transmitted and reflected streamers can be observed on shorter timescales, here 850 ns. As detailed in section II.3, the present fast ICCD characterization has been achieved using $L_{kin}$ = 850 scans with a delay ($\delta$) incremented every 1 ns.

The Figure 12 is divided in 5 subfigures with d = 1 cm (Figure 12a), d = 3 cm (Figure 12b), d= 5 cm (Figure 12c), d= 7 cm (Figure 12d) and d = 9 cm (Figure 12e). In each of these subfigures, the integrated emission of the guided streamers is plotted versus time, considering different $x_k$ locations along the x axis. Each of these profiles contains at least one main peak associated with the most emissive part of the streamer's head, typically the ionization front. This main peak is considered as a reference both for locating the streamer within the capillary and for normalizing the entire integrated emission profile between 0 and 1.

For d = 1 cm, all the integrated emission profiles show two types of guided streamers: the transmitted guided streamer ($GS_t$) after its passage through GAEL and the counter-propagation of a guided streamer ($GS_r$) which is reflected by GSEL. $GS_t$ appears at t = 30 ns with the highest intensity (here normalized to 1) and shifts towards the increasing values of x. On the contrary, $GS_r$ has a much lower intensity and counter-propagates. Unsurprisingly, the time interval separating these two peaks narrows with increasing values of x. By correlating these integrated emission profiles with the current intensity profiles of Figure 10, it turns out that the first and second emission peaks correspond to the $\zeta_t^+$ and $\zeta_r^-$ conductive peaks respectively.

For d = 3, 5, 7 and 9 cm, the integrated emission profiles can be performed for $x < x_{GAEL}^-$ and $x > x_{GAEL}^+$. In that latter case, the profiles are similar to those obtained for d = 1 cm, i.e. a first peak corresponding to ($GS_t$, $\zeta_t^+$) and the second one to ($GS_r$, $\zeta_r^-$). The influence of d on the time separating these two peaks is quite clear and discussed in the Section IV. Interestingly, the integrated emission profiles measured for $x < x_{GAEL}$ reveal not only a single peak associated to $GS_i$ but at least one other peak that evidence the existence of at least one reflected guided streamer. In Figure 12b (d = 3 cm), the peak corresponding to the incident guided streamer ($GS_i$) shows an intensity of 1, followed by two smaller peaks corresponding to reflected guided streamers: $GS_{r'}$ (closer to $GS_i$ with an intensity of 0.5-0.7) and $GS_{r''}$ (further away from $GS_i$ with a relative intensity of 0.2-0.3). While $GS_{r''}$ is only visible for d = 3 cm, $GS_{r'}$ is still clearly detected for the higher values of d in Figure 12c, 12d and 12e. Besides, the more d increases, the further $GS_{r'}$ can move away from $GS_i$.

The peak associated to $GS_{r'}$ should not be confused with a $GS_r$ peak, the latter one resulting from a reflection on GSEL. The peak associated to $GS_{r'}$ is measured at an instant that comes much earlier before $GS_r$ is generated. As an example, in Figure 12b, $GS_i$ appears at t = 5 ns , is transmitted at t = 70 ns and is reflected by GSEL at t ≈ 450ns. Much earlier, the $GS_{r'}$ and $GS_{r''}$ peaks have appeared at 50 and 85 ns respectively. In our experimental setup, $GS_{r'}$ and $GS_{r^*}$ can only be detected by fast ICCD imaging. Electrical analysis would only be possible by placing another current monitor between the HV electrode and GAEL.

The Figures 12a, 12b, 12c, 12d and 12e are also plotted as contour plots in Appendix IX.2.

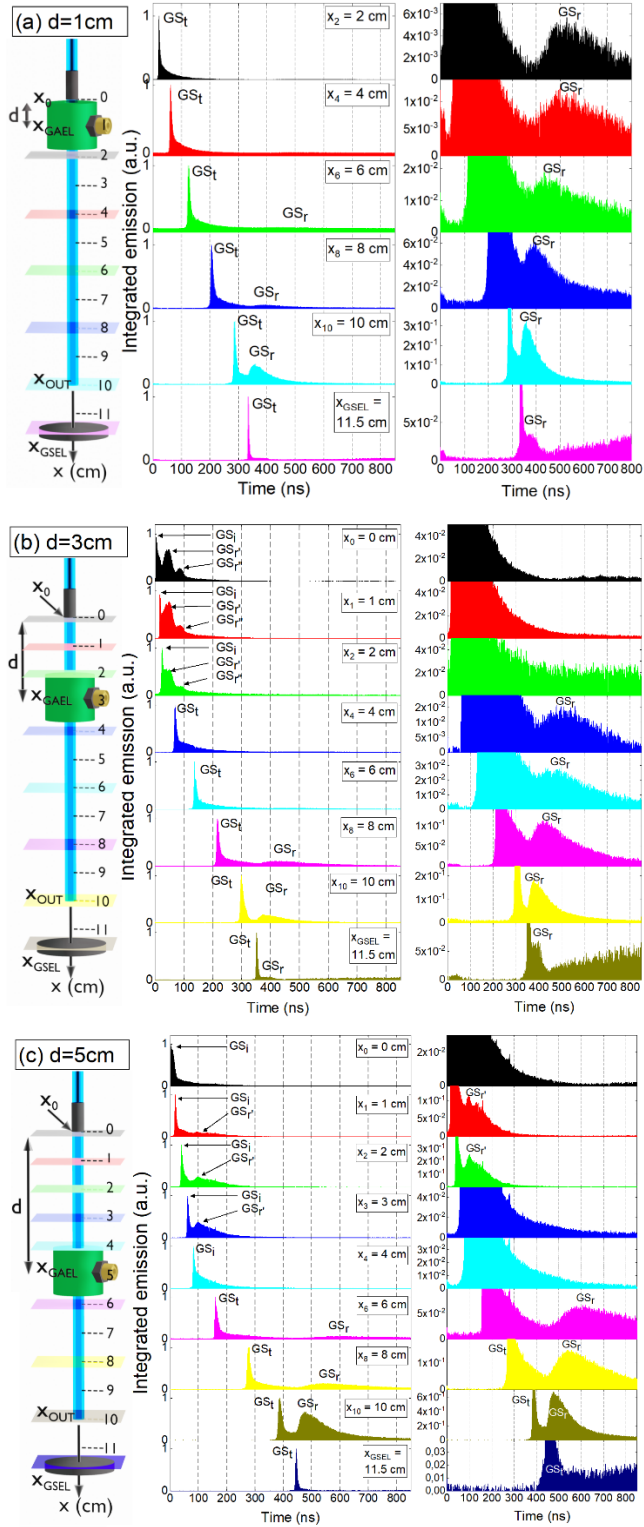

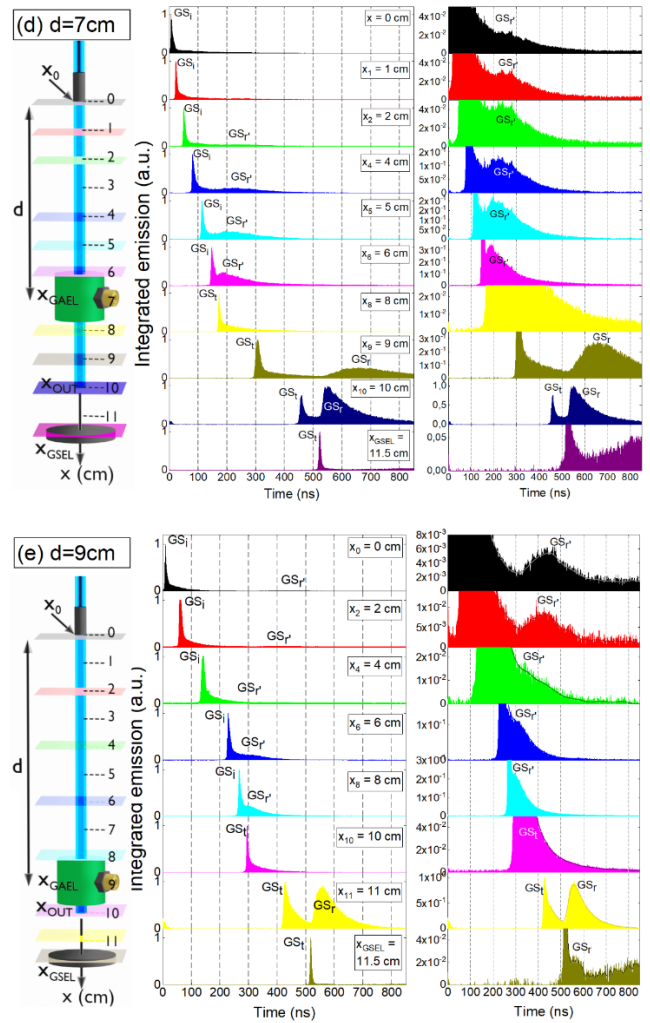

*Figure 12. Integrated emission profiles of the guided streamers propagating in the plasma gun device and interacting with the two distant grounded electrodes: the grounded annular electrode (GAEL) and the grounded surface electrode GSEL). (a) d = 1 cm, (b) d = 3 cm, (c) d = 5 cm, (d) d = 7 cm, (e) d = 9 cm. For each subfigure, a simplified assembly diagram is represented on left part while the integrated emission profiles are magnified in the right part.*

### III.2.4. Relevant parameters obtained from ICCD characterization

Several parameters can be extracted from the previous integrated emission profiles, namely the characteristic propagation times (as already achieved from electrical study in Figure 10), the ratios of the integrated emission peaks and the velocity of the guided streamers in the immediate vicinity of GAEL and GSEL.

While electrical characterizations (Sections III.2.1. and III.2.2.) only allow the measurement of characteristic propagation times between $CM_1$ (GAEL) and $CM_2$ (GSEL), fast ICCD imaging allows to accurately localize the guided streamers at any x coordinate of the transparent capillary except in the region hidden by $CM_1$ which is 16 mm thick. From Figure 12a to Figure 12e, the instant at which the head of a guided streamer occupies the $x_{GAEL}$ position can be approximated by $t = \frac{t(x_{GS_r} = x_{GAEL}^-) - t(x_{GS_i} = x_{GAEL}^+)}{2}$. The Figure 13a indicates the durations required by the guided streamers to propagate from $x_k$ to $x_{GSEL}$ (filled symbols) as well as to counter-propagate from $x_{GAEL}$ to $x_k$ (open symbols). As an example, for d = 5 cm, a guided streamer at the $x_2$ location requires 375 ns to reach GSEL while in the case of a counter-propagation from GSEL to reach the $x_6$ location, 130 ns are expected (green curve). Besides, it appears that the forward propagation times are always longer than the backward ones. As an example, for d = 1 cm and x = 2 cm, a duration of 300 ns is required for forward propagation versus only 200 ns for backward propagation. This figure permits also to distinguish the cases where $x_k \neq d$ (circle symbols) and where $x_k = d$ (square symbols). This latter case corresponds to the forward/backward propagation ways between CM$_1$ and CM$_2$: they can be used to determine the characteristic propagation times as previously achieved in the electrical characterization. Hence, $\tau_f$ is obtained by measuring the difference between the instant at which GS$_t$ reaches GSEL and the instant at which GS$_t$ (or GS$_i$) was located at GAEL. Similarly, $\tau_b$ is the difference between the instant when GS$_r$ reaches GAEL and the instant when GS$_t$ reached GSEL. These particular values are plotted versus d in Figure 13b which indicates a decrease in $\tau_f$ and $\tau_b$ as a function of d. This illustrates that the further GAEL is from the HV electrode, the faster $GS_t$ propagates through the capillary. Thus, while $\tau_f$ drops from 310 ns (d = 1 cm) to 250 ns (d = 9 cm), $\tau_b$ is generally smaller since it varies from 180 ns (d = 1 cm) at 70 ns (d = 9 cm).

From Figure 12, it is possible to measure the integrated emission of each peak, whether for the incident guided streamer ($I_i$), the transmitted guided streamer ($I_t$) or the reflected guided streamers ($I_r$ and $I_{r'}$). Then, their values can be correlated by plotting the $I_{r'}/I_i$ ratio (Figure 14a) and the $I_r/I_t$ ratio (Figure 14b) on a log scale as a function of x and for different values of d. For a given position of GAEL (d constant) in Figure 14a, the $I_{r'}/I_i$ ratio becomes larger for increasing values of x because the reflected guided streamer is detected closer to GAEL, i.e. no more than 1 cm away. As an example with d = 7 cm, $\left(\frac{I_{r'}}{I_i}\right)_{x=2cm} = 8 \times 10^{-3}$ while $\left(\frac{I_{r'}}{I_i}\right)_{x=8cm} = 2 \times 10^{-1}$. The same behavior is observed in Figure 14b with d = 3 cm for example: $\left(\frac{I_r}{I_t}\right)_{x=8cm} = 1.1 \times 10^{-1}$ while $\left(\frac{I_r}{I_t}\right)_{x=10cm} = 5.4 \times 10^{-1}$. This later observation is consistent with the works of Darny et al. where the reflected guided streamers always present a higher magnitude than that of the transmitted guided streamers [43]. Besides, for a given $x_k$ coordinate, a decrease in d drives to higher $I_{r'}/I_i$ ratios but lower $I_r/I_t$ ratios.

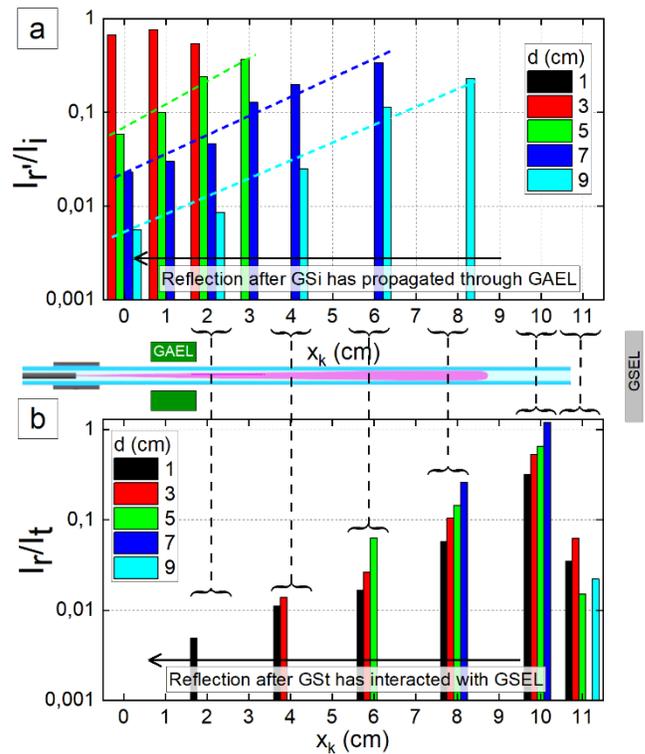

Figure 14. (a) Emission ratio of the incident/reflected guided streamers interacting with GAEL, (b) Emission ratio of the reflected/transmitted guided streamers interacting with GSEL.

In addition to the characteristic propagation times in Figure 13 and to the emission ratios in Figure 14, a third parameter of interest is the velocity of the streamers before/after interacting each of the two distant grounded electrodes. To evaluate the values of these velocities, we have first plotted in Figure 15a the profiles of the guided streamers in a time-space diagram, for different values of d. As sketched in inset and as verified for any experimental profile, an incident guided streamer splits into a first reflected (r') guided streamer, eventually a second reflected (r'') streamer that is not represented here for the sake of clarity and a transmitted (t) guided streamer that appears after passing through GAEL. Then, this guided streamer impinges with GSEL, driving to an additional relaxion (r). Each profile is composed of experimental data (solid line) and interpolated values (dashed line) when the streamer (counter)propagates through GAEL. All the profiles, whether forward or backward are non-linear, thus evidencing the existence of acceleration and deceleration regions. As an example, an acceleration region of $GS_t$ is clearly visible in the 15mm-air gap separating the capillary's outlet from GSEL ($10.0\ cm < x < 11.5\ cm$). Besides, the location of GAEL has a strong influence on the guided streamers' dynamics. As an example, the duration required by the guided streamers to bridge the HV electrode to GSEL is only 335 ns for d = 1 cm and becomes as high as 530 ns for d = 9 cm. From this figure, it is possible to measure the spatio-temporal coordinates of the guided streamers (i) in the immediate vicinity of GAEL, at $x_{GAEL}^-$ and $x_{GAEL}^+$ and (ii) at the surface of GSEL at $x_{GSEL}$, as sketched in Figure 3. As a result, the velocities of the incident guided streamer ($v_{GSi}$) at $x_{GAEL}^-$ and of the transmitted guided streamer ($v_{GSt}$) at $x_{GAEL}^+$ can be plotted as a function of d in Figure 15b. Similarly, the velocities of the transmitted ($v_{GSt}$) and reflected ($v_{GSr}$) guided streamers interacting with the grounded metal target are plotted versus d in Figure 15c.

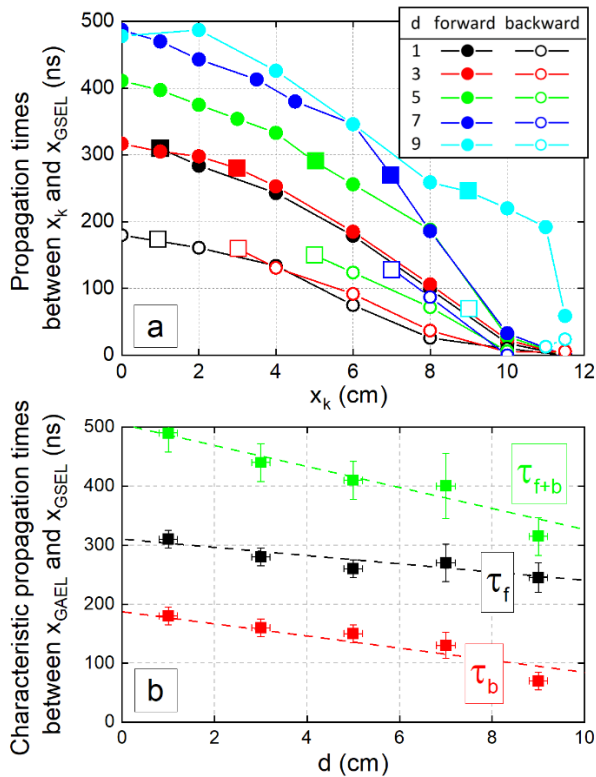

Figure 13. (a) Propagation times required by guided streamers to reach $x_{GAEL}$ from $x_k$ locations (forward, filled symbols) or to reach specific $x_k$ locations from $x_{GAEL}$ (backward, open symbols). The square symbols indicate the condition $x_k = d$, (b) Forward, backward and round-trip propagation times measured as a function of d, i.e. between GAEL (fixed location) and GSEL (variable location).

In Figure 15b, the velocity of the guided streamers is always reduced after passing through GAEL, e.g. decreasing from $v_{GSi} = 960\ km/s$ to $v_{GSt} = 300\ km/s$ at d = 3 cm. Interestingly, this trend is even more pronounced for the increasing values of d, so that the value of $v_{GSt}$ can change by 50 %. On the contrary, such trends are not observed when $GS_t$ interacts with GSEL (grounded surface electrode): as shown in Figure 15c, the velocities are always significantly higher for $GS_r$. Hence, when GAEL is placed 3 cm away from the HV electrode, $GS_t$ impinges the grounded metal target at $v_{GSt} = 400\ km/s$ while its reflected streamer leaves GSEL at a higher velocity: $v_{GSr} = 580\ km/s$. It is also worth stressing that the reflections velocities are much higher for streamers interacting with GAEL than GSEL. Indeed, if we consider the condition d = 5 cm, it turns out that $v_{GSr} = 1400\ km/s$ while $v_{GSr} = 600\ km/s$.

# IV. Discussion

## IV.1. Complementarity of electrical and optical analysis

### IV.1.1. Characteristic propagation times

In this study, we have demonstrated how to detect reflected and transmitted guided streamers combining two analysis techniques that rely on fundamentally different physical principles: the first measures the electrical properties of guided streamers and the other their optical properties. To assess this complementarity, Figure 16 shows the characteristic propagation times obtained by electrical analysis (Figure 11c) and by optical analysis (Figure 13b) for several values of d. If the trends remain the same, it appears that a quasi-constant temporal gap is measured, of the order of 50-100 ns.

One reason that could have accounted for this delay is the thermal drift of not only the voltage generator but also the ICCD camera between the times they are turned on and their period of use. To overcome this artifact, we always let the cold plasma jet run for 15 minutes before launching the acquisitions by fast ICCD imaging (which always remained in operation, thus in thermal equilibrium). This same duration of 15 minutes was respected before launching any measurement, again to allow both the voltage generator and the plasma source to reach this same thermal equilibrium.

Therefore, two other assumptions have been proposed to explain this 50-100 ns discrepancy. First, this discrepancy could result from a different number of acquisitions between (i) the electrical approach whose statistic relies on a triplicate of 3 oscillograms, i.e. 3 acquisitions and (ii) the optical approach whose statistic is performed on a triplicate of 3 trains of guided streamers, each train containing $N_{GS} = 25\ 000$ guided streamers (Figure 6). Second, the values obtained by electrical characterization could be slightly underestimated due to the use of the Butterworth low-pass filter (Section II.2.3.). Despite these slight discrepancies, the trends remain the same and consolidate our conclusions: (i) the reflection of the guided streamers is always faster than that of the transmitted streamers, (ii) the closer GAEL is to GSEL, the shorter $\tau_b$ is.

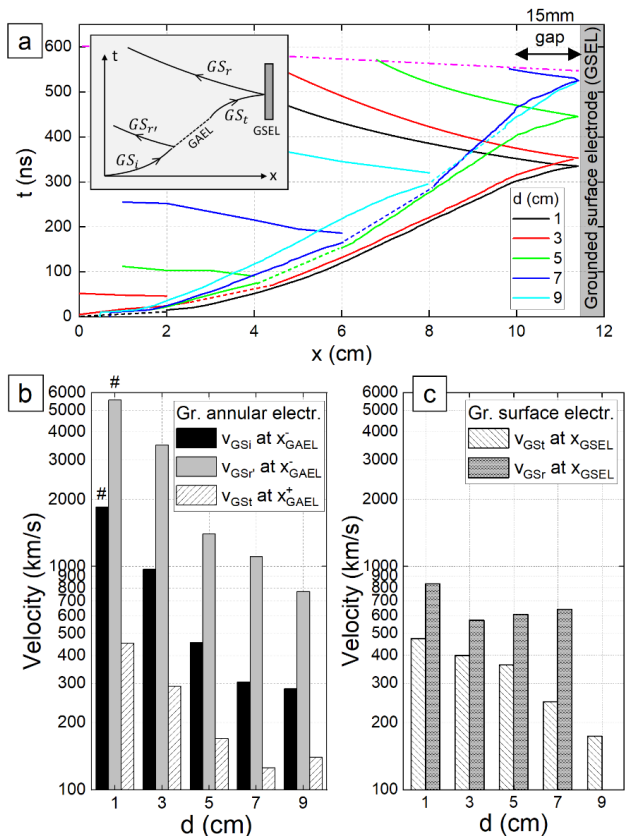

***Figure 15. (a) Spatio-temporal diagram of the incident, reflected and transmitted guided streamers (b) Velocity of the guided streamers before/after passing through GAEL (grounded annular electrode). (c) Velocity of the guided streamers before/after interacting with GSEL (grounded surface electrode). All the curves are plotted for d = 1, 3, 5, 7 and 9 cm. (#: estimation by extrapolation).***

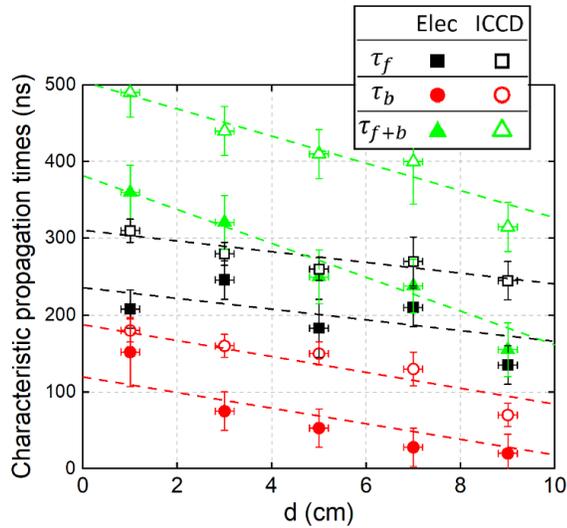

**Figure 16. Characteristic propagation time as a function of d measured using the electrical or optical approach.**

### IV.1.2. Electrical charge of transmitted and reflected streamers

Fast ICCD imaging alone cannot provide information about the electrical charge carried by the heads of guided streamers; it just allows to decipher whether these guided streamers are incident, transmitted or reflected (Figure 7, Figure 8). Likewise, the electrical analysis alone allows only to identify current peaks (either capacitive or conductive) when guided streamers pass through the current monitors, at only two positions in our experimental set-up. This being said, electrical characterization has a major advantage that deserves to be emphasized: it allows us to identify the polarization (positive/negative) of the conductive current peaks associated with the streamers and thus the overall electrical charge they carry in their heads. Thus, by combining the electrical and optical approaches, a conductive current peak $\zeta_t^+$ associated with a transmitted guided streamer $GS_t$ can be written as: $GS_t^+$. Similarly, a conductive current peak $\zeta_r^-$ associated with a reflected guided streamer $GS_r$ can be written as: $GS_r^- -$.

### IV.1.3. Highlighting multiple reflection phenomena

While the electrical analysis alone only shows one type of reflection (the r reflection corresponding to the interaction of $GS_t^+$ on the grounded metal target), the fast ICCD imaging reveals 4 types of reflection:

- r: reflection corresponding to the counter-propagation of a negative guided streamer initiated by $GS_t^+$ at the capillary's outlet. This reflection is the only one to be detected both optically and electrically (Figure 10). Our experimental results are in agreement with the works of Babaeva et al. modelling the counter-propagation of a streamer approaching and reflecting a metal grounded target [34].
- R: reflection corresponding to the counter-propagation of a guided streamer initiated from GSEL and appearing after $GS_r^-$. The R reflection remains confined in the 15 mm gap and cannot reach the outlet's capillary. If the polarity of this streamer cannot be measured by electrical analysis, it can reasonably be assumed to be negative, as previously demonstrated with the counter-propagation of guided streamers with a negative charge. If so, the following notation can be used: $GS_R^-$.

- r': reflection corresponding to the counter-propagation of a guided streamer initiated from the hollow region of GAEL ($GS_{r'}^-$). Its propagation velocity is very high and is all the greater the closer GAEL is to the HV electrode. Given our experimental setup, this type of reflection can only be demonstrated by fast ICCD imaging. However, its existence could also be demonstrated by electrical analysis by placing a third current monitor between the HV electrode and GAEL (see Figure 3). Here, since $GS_{r'}$ is only detected by fast ICCD imaging, the sign of its electrical charge cannot be deduced. However, it is assumed to be negative, hence the notation $GS_{r'}^-$. This hypothesis relies on the previous result (r reflection on GSEL). Besides, Figure 15 indicates that the reflected streamers are always faster than the incident/transmitted ones: $v_{GS_{r'}^-} > v_{GS_t^+}$ but also $v_{GS_{r^-}} > v_{GS_t^+}$. This result may appear in contradiction with previous research works where negative guided streamers are slower than the positive ones [44], [45]. However, in our case the reflections result from the interaction with a grounded electrode and in a counter-propagation configuration, i.e. in the ionic trace of a previous guided streamer.
- r'': reflection corresponding to the counter-propagation of a guided streamer initiated from GAEL and appearing after $GS_{r'}$. The existence of this streamer can only be demonstrated by fast ICCD imaging and under the condition d = 3 cm. Demonstrating its existence at d = 1 cm since is impossible since it is hidden by GAEL which is 16 mm thick. For d = 5 cm, $GS_{r'}$, no longer exists or has been considerably attenuated; this is the case for $GS_{r'}$, whose integrated emission is strongly reduced for d ranging from 3 to 5 cm (Figure 12c). The study of $GS_{r''}$ would deserve a dedicated study to understand the mechanisms underlying its generation.

## IV.2. Propagation mechanisms & Equivalent electrical model

### IV.2.1. Electron sources

In this study, the experimental device has two distant ground electrodes (GAEL and GSEL), whose electrical potential is always 0 V. It should be remembered that an electrical ground corresponds to a reservoir containing an infinite number of free electrons and whose electrical potential is always zero Volt. Thus, if a flow of positive charges (total charge Q+) is transferred to the ground, then it generates an equivalent flow of electrons (total charge Q−) to maintain its zero potential ($V_{ground}$ = 0 V). In our experimental setup, the distant grounded electrodes are not a mandatory for the generation and propagation of guided streamers. Whether it is positive or negative, a guided streamer can propagate from one electrode to another through photo-ionization processes that take place in its head to generate electrons. In the case of a positive guided streamer, the electron density obtained is lower than that of the positive ions produced, so that the overall charge of the guided streamer's head is positive (Q+). When this streamer comes into contact with a grounded electrode, the latter can release a certain number of electrons (global charge Q−) in streamer's tail.

### IV.2.2. Propagation from the HV electrode to GAEL

As it propagates, $GS_i^+$ leaves behind an ionic trace in the volume of the capillary as well as polarization on the inner walls. This surface polarization is sketched in Figure 17a which introduces an equivalent electrical model of the guided streamers. During its passage through GAEL, i.e. from $x_{GAEL}^-$ to $x_{GAEL}^+$ (Figure 3), this surface polarization is considerably reduced (or even cancelled). The electric charge carried by $GS_i^+$ splits in two components: (i) $Q_t^+$ carried by the guided streamer transmitted by GAEL and (ii) $Q_{leak}^+$ which leaves the capillary as a capacitive leakage current $I_\kappa = \frac{dQ_{leak}^+}{dt}$ by passing through which stand for the capacitances of the capillary thickness and the air gap (separating the capillary from GAEL) respectively. Since the streamers are generated at atmospheric pressure and the excited/ionized particles (electrons, positive ions) are governed by energy distribution functions, it is assumed that GAEL locally induces a potential barrier that separates the most energetic particles (particles that can pass through GAEL and thus constitute the transmitted streamer of charge $Q_t^+$) from the least energetic particles (particles that cannot pass through GAEL and that are dissipated in the form of a capacitive leakage current involving $Q_{leak}^+$). To maintain its potential at 0V, GAEL must provide a flow of electrons (of overall charge $Q_{GAEL}^-$) that returns into the capillary, or more precisely into the ionic tail of $GS_i^+$. Then, $Q_{GAEL}^-$ gives rise to a negative space charge zone characterized by an electron density much higher than the ionic density. In the model of Figure 17, the open/closed states of the $K_1(t)$ and $K_2(t)$ switches must reverse, so that electrons can move to the increasing potentials, allowing this negative space charge region to counter-propagate towards the HV electrode as a negative guided streamer ($GS_{r1}^-$).

### IV.2.3. Propagation from GAEL to GSEL]

When $GS_t^+$ propagates from GAEL to GSEL, it first interacts with the inner walls of the dielectric capillary, leaving behind an ionic trace in the volume of the capillary and a positive polarization of the inner walls, as shown in Figure 17b. Then, the head of $GS_t^+$ is separated from GSEL by a layer of air that can be modeled by $C_{air}^*(x,t)$. As the streamer approaches GSEL, the air thickness decreases, thus increasing the value of this capacitance. When $GS_t^+$ reaches GSEL, the capacitance becomes infinite and therefore behaves as a simple short circuit. $GS_t^+$ directly impinges on GSEL to transfer its charge $Q^+$. Still to respect the condition $V_{ground} = 0V$, GSEL returns a flow of electrons, of overall charge $Q_{GSEL}^-$ and of very high mobility. Unlike the case where the interaction operates with GAEL, we assume here that the totality of $Q^+$ is transferred to GSEL.

### IV.2.4. Propagation from GSEL to the HV electrode]

The negative space charge region resulting from the transfer of $Q_{GSEL}^-$ into the 15 mm gap gives rise to a negative guided streamer ($GS_r^-$) which can then counter-propagate towards GAEL and then the HV electrode. As $GS_r^-$ counter-propagates, the value of $C_B(x,t)$ increases.

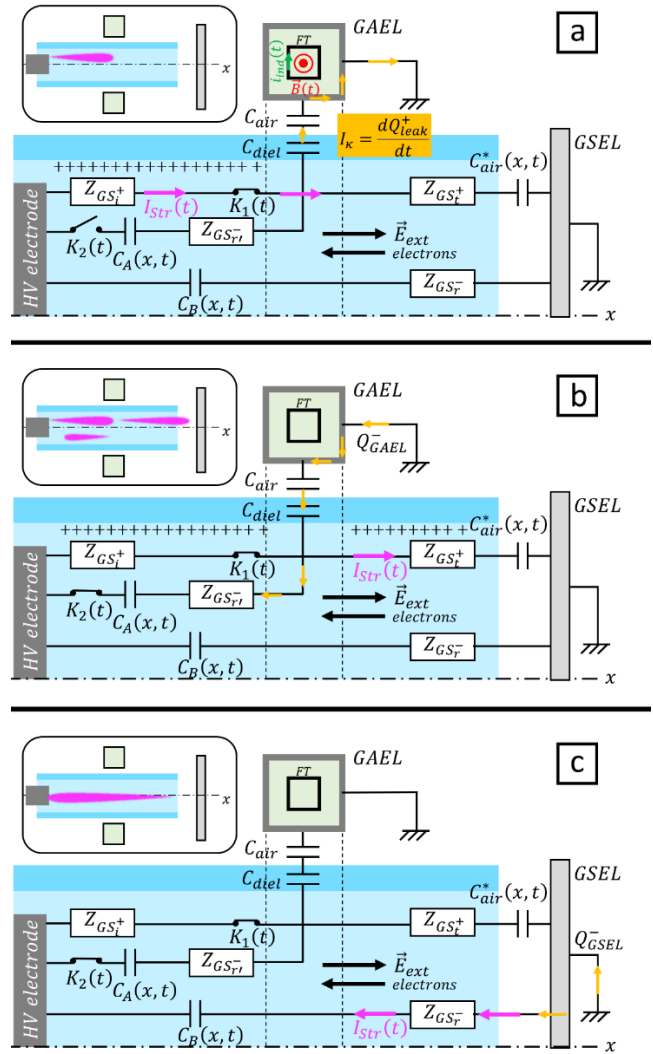

**Figure 17.** Equivalent electrical model explaining the (counter)propagation of guided streamers in a plasma gun device interacting with two distant grounded electrodes (GAEL and GSEL).

## IV.3. Reflected guided streamer versus guided return stroke

As sketched in Figure 18, a streamer is an ionization wave that propagates longitudinally and that transports electrical charges as well as radiative species over long distances. Thanks to the ionizing mechanisms directly generated in its pre-head region (e.g. photo-ionization), a streamer can propagate in a gaseous environment even if it does not contain any charged particle. Hence, in the case of a positive guided streamer propagating along a capillary from the anode to the cathode, two important characteristics are noteworthy: (i) the positive charges of the streamer are left on the inner walls during propagation and (ii) a second wave is generated after the positive streamer has interacted with the anode. Then, $CM_1$ (i.e. GAEL) can show the same positive current peak following the passage of the second wave depending on whether it has a negative charge as it propagates backwards (reflected guided streamer) or a positive charge as it propagates forwards (return stroke) [46].

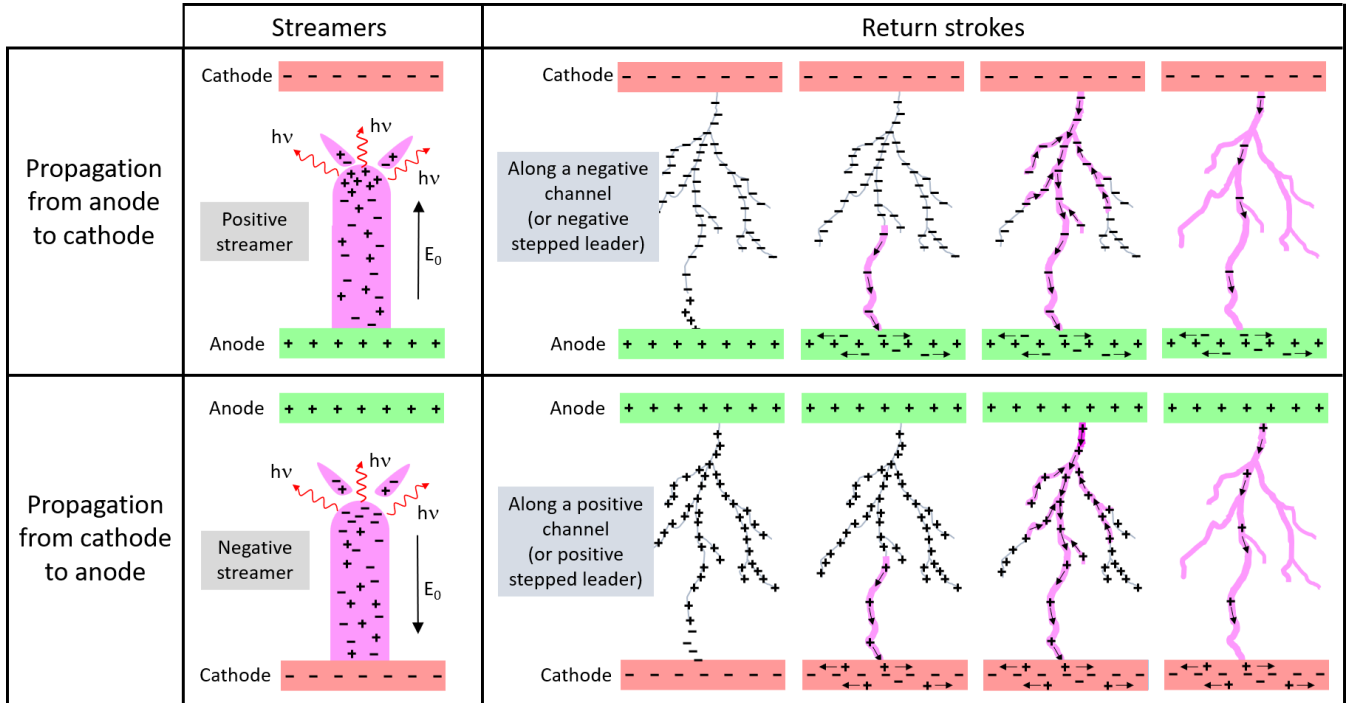

*Figure 18. Synoptic diagram explaining the propagation mechanisms of streamers (whether positive or negative) and return strokes (whether along a negative or positive channel).*

In the case of the **reflected guided streamer**, the counter-propagation results from a local electric field induced by two space charge regions: the negative charged region of the reflected wave and the residual positive charges that line the inner walls of the capillary. This mechanism alone would be sufficient to generate the counter-propagation, hence making photo-ionization no more a mandatory.

In the case of the **return stroke**, the propagation mechanics is different, as sketched in Figure 18. Return stroke corresponds to the most luminous and noticeable part of a lightning discharge [47]. Once a conductive channel bridges the air gap between a negative charge excess in the cloud (cathode) and the positive surface charge excess on the ground (anode), a large drop in resistance is observed across the lightning channel. In the case of a negative channel (or negative stepped leader), electrons accelerate rapidly as a result in a zone beginning at the point of attachment, which expands across the entire leader network at up to one third of the speed of light [48].

In atmospheric physics, return strokes are therefore non-guided ionization waves which require negative charge channels to counter-propagate. In our experimental setup, the situation is quite different since the ionization wave is guided and the conductive channel is composed of positive charges. This makes a main difference with a conventional return stroke, although the existence of (guided) return strokes propagating in positively charged channels may also be considered (Figure 18) [49]. In that latter case, CM$_1$ would measure a forward flow of positive charges (instead of a backward flow of negative charges) with a velocity expected to be close to that of the first positive guided streamer.

However, Figure 11c states exactly the opposite with $\tau_b < \tau_f$, i.e. the backward wave is faster than the forward one. For this reason, the assumption of a reflected negative guided streamer has appeared more relevant than the guided return stroke.

# V. Conclusion

A DC high-voltage power supply has been utilized to generate trains of positive guided streamers in a plasma gun device. Its dielectric capillary, supplied in helium, interacts with two distant electrodes: a grounded annular electrode (coaxially centered around the capillary) and a grounded surface electrode (15 mm away from capillary's outlet). By combining electrical analysis and fast ICCD imaging, we have developed a methodological approach that allowed to demonstrate novel results:

- Guided streamers can be reflected by passing through the air gap of a grounded annular electrode, without any kind of impact on solid-state target.

- Reflected guided streamers carry a negative charge. Although predicted by theory, this result is now demonstrated using current monitor CM1 and using Ampère's right-hand grip rule.

- Four types of reflections have been identified: Two reflections following an impact with the grounded metal surface: one with sufficiently high kinetic energy to counter-propagate over long distances and enter the capillary as a negative guided streamer I, and the other with lower kinetic energy, so that the reflected negative streamer I can only counter-propagate in the 15 mm gap, without being able to penetrate the capillary. In addition, two other reflections have been evidenced once the incident guided streamer has interacted with the grounded annular electrode. These r' and r'' reflected guided streamers are only detectable by fast ICCD imaging and their electrical charge is assumed to be negative.

- Streamers propagating backward (reflection) are faster than those propagating forward (incident or transmitted), especially after a reflection involving GAEL (rather than GSEL). Velocities as high as 3000 km/s are thus obtained for d = 3 cm (Figure 15b).

- GAEL is always located between the HV electrode and GSEL. Its relative position has a significant influence on the optical emission of the guided streamers but also on their characteristic propagation times. Hence, bringing GAEL closer to GSEL, contributes to significantly reduce $\tau_f$ (typically from 160 ns to 50 ns), as well as $\tau_b$ (typically from 250 ns to 200 ns) and therefore $\tau_{f+b}$ (typically from 420 ns to 250 ns).

- Whatever the type of distant grounded electrode, the amplitude of the reflected streamers decreases exponentially. For a given position ($x_k$), the more GAEL is close to the HV electrode, the more the amplitude of the streamers reflected by GSEL decreases (Figure 14b) while the amplitude of the streamers reflected by GAEL increases (Figure 14a).

Based on these results, an equivalent electrical model is proposed to better understand guided streamers dynamics. In this model, the grounded electrodes are defined as reservoirs containing an infinite number of free electrons, releasable at any time to always maintain a 0 Volt potential, especially when GAEL or GSEL is exposed to a flow of positive charges from the incident guided streamer.

Although this experimental work has allowed to investigate the physics of guided streamer propagation in a purely fundamental framework, it could have strong spin-offs in applied research. For example, in plasma medicine, the innovation of therapeutic plasma sources could require on-board sensors such as current monitors to measure the currents in real time during the patient therapy.

# VI. Acknowledgements

The authors would like to thank Sorbonne Université and the Île-de-France Region for supporting fundamental research in plasma physics by co-funding the $^{PF2}$ABIOMEDE platform project (Sesame 2016).

# VII. Data Access statement

The data that support the findings of this study are available upon reasonable request from the authors.

# IX. Appendix

## IX.1. Analytical description of the characteristic propagation times

From the electrical current peaks from Figure 9 and Figure 10, five characteristic durations can be defined:

- $\tau_{\kappa\kappa}$ is the duration between the instant when the capacitive current peak is measured by CM$_1$ and the instant when the capacitive current peak is measured by CM$_2$, i.e. respectively the black and red $\kappa$ peaks in Figure 9b.

- $\tau_{\kappa\zeta}$ is the duration between the conductive current peak of a guided streamer (either $\zeta_i^+$ or $\zeta_t^+$) and its related capacitive current peak (either $\kappa_i^+$ or $\kappa_t^+$).

- $\tau_f$ is the time required by $GS_t$ to propagate forward (f) from CM$_1$ to GSEL, which is roughly the same as from CM$_1$ to CM$_2$. In consequence, $\tau_f$ is the duration between $t\left(x_{\zeta_i^+} = x_{GAEL}\right)$ and $t\left(x_{\zeta_t^+} = x_{GSEL}\right)$ which corresponds respectively to the instant at which the conduction current peak of $GS_i$, passes through CM$_1$ (and therefore GAEL) and the instant at which the conduction current peak of $GS_t$ reaches GSEL (see equation {8a}).

- $\tau_b$ is the duration required by $GS_r$ to propagate backward (b) or counter-propagate from GSEL to GAEL, which is roughly the same as from CM$_2$ to GAEL. In consequence, $\tau_b$ is the duration between $t\left(x_{\zeta_r^-} = x_{GAEL}\right)$ and $t\left(x_{\zeta_t^+} = x_{GSEL}\right)$ which correspond respectively to the conduction current peak of $GS_r$ when it passes through CM$_1$ and the instant at which the conduction current peak of $GS_t$ is measured at GSEL (see equation {9a}).

- $\tau_{f+b}$ corresponds to the time required by a guided streamer to propagate (forward, f) from GAEL to GSEL and then come back to GAEL (backward, b), that is to say the time required to achieve the GAEL-GSEL-GAEL round trip. In consequence, $\tau_{f+b}$ is the duration between $t\left(x_{\zeta_t^+} = x_{GAEL}\right)$ and $t\left(x_{\zeta_r^-} = x_{GAEL}\right)$ which correspond respectively to the instant when $GS_i$ – and therefore its related conduction current peak ($\zeta_t^+$ in Figure 9b) – passes through GAEL and the instant when $GS_r$ counter-propagates so that its conduction peak ($\zeta_r^-$ peak in Figure 9b) reaches GAEL (see equation {10a}).

- Equations {8b}, {9b} and {10b} are a generalization of equations {8a}, {9a} and {10a} and for which the characteristic propagation times are not referenced to x$_{GAEL}$. These equations will be particularly useful in the fast ICCD imaging analysis of Figure 13.

| Characteristic propagations times between the 2 distant grounded electrodes | Characteristic propagations times including any position x$_k$ |
|---|---|

| | | | |
|---|---|---|---|
| $\tau_f$ $= t(x_{\zeta_t^+} = x_{GSEL})$ $- t\left(x_{\zeta_l^+} = x_{GAEL}\right)$ | {8a} | $\tilde{\tau}_f(x)$ $= t(x_{\zeta_t^+} = x_{GSEL})$ $- t\left(x_{\zeta_l^+} = x_k\right)$ | {8b} |
| $\tau_b = t_{GAEL}^{\zeta_r^-} - t_{GSEL}^{\zeta_t^+}$ | {9a} | $\tilde{\tau}_b(x)$ $= t(x_{\zeta_r^-} = x_k)$ $- t(x_{\zeta_t^+} = x_{GSEL})$ | {9b} |
| $\tau_{f+b} = t_{GAEL}^{\zeta_r^-} - t_{GAEL}^{\zeta_t^+}$ | {10a} | $\tilde{\tau}_{f+b}(x)$ $= t(x_{\zeta_r^-} = x_k)$ $- t\left(x_{\zeta_t^+} = x_k\right)$ | {10b} |

**Table 1.** *Characteristic propagation times either to bridge the two distant grounded electrodes (GAEL and GSEL) or to bridge any x positions. $\tilde{\tau}_f$ : forward propagation time, $\tilde{\tau}_b$: backward propagation time, $\tilde{\tau}_{f+b}$:round-trip propagation time*

## IX.2. Integrated emission profiles of the guided streamers represented as contour plots

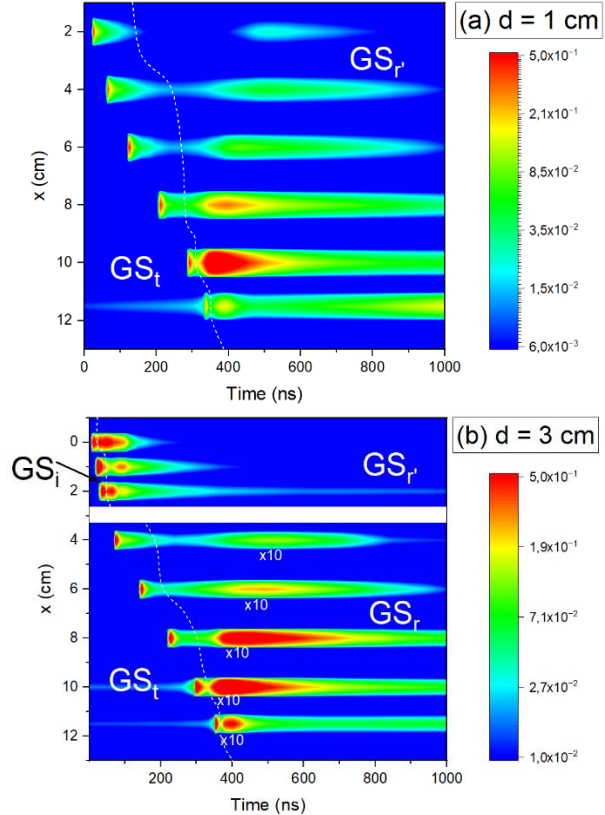

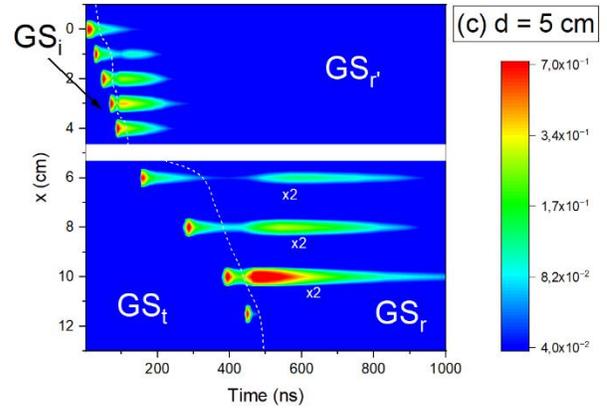

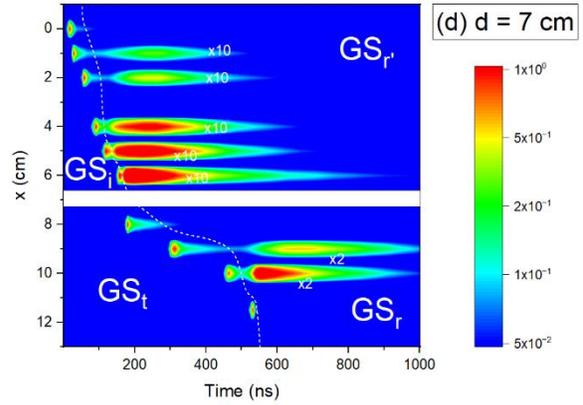

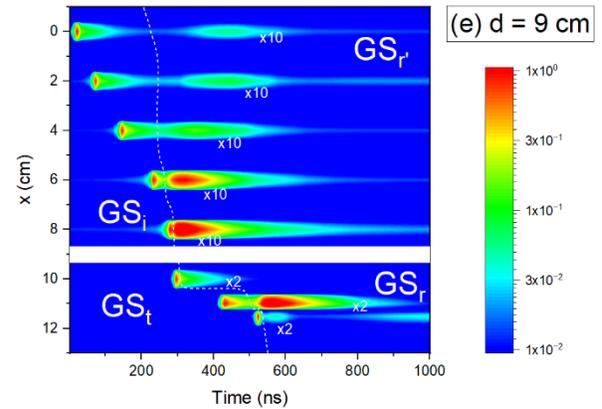

**Figure 1.** *Integrated emission profiles of the guided streamers represented as contour plots for (a) d = 1 cm, (b) d = 3 cm, (c) d = 5 cm, (d) d = 7 cm, (e) d= 9 cm. The emission intensity of the reflected guided streamers can be multiplied by a coefficient (x2 or x10) so as to be visible in the subfigures.*